\documentclass[12pt]{article}                  
\usepackage{osajnl}
\usepackage{graphicx}
\begin{document}
\title{Systematic effects induced by a \\ flat isotropic dielectric slab}
\author{Claudio Macculi$^{*,}$\footnote{present address: INAF-IASF Roma, Via del Fosso del Cavaliere 100, 00133 Roma, Italy},
        Mario Zannoni$^{\sharp}$, Oscar Antonio Peverini$^{\ddag}$,
        Ettore Carretti$^{*,}$\footnote{present address: INAF-IRA Bologna, Via Gobetti 101, 40129 Bologna, Italy},
        Riccardo Tascone$^{\ddag}$, Stefano Cortiglioni$^{*}$}
\affiliation{$^{*}$INAF-IASF Bologna, Via Gobetti 101, 40129
Bologna, Italy
\\
$^{\sharp}$Dip. di Fisica, Univ. di Milano - Bicocca, P.zza della
Scienza 3, 20126 Milano, Italy
\\
$^{\ddag}$CNR-IEIIT, C.so Duca degli Abruzzi 24, 10129 Torino,
Italy}
\begin{abstract}
The instrumental polarization induced by a flat isotropic
dielectric slab in microwave frequencies is faced. We find that,
in spite of its isotropic nature, such a dielectric can produce
spurious polarization either by transmitting incoming anisotropic
diffuse radiation or emitting when it is thermally inhomogeneous.
We present evaluations of instrumental polarization generated by
materials usually adopted in Radioastronomy, by using the Mueller
matrix formalism. As an application, results for different slabs
in front of a 32 GHz receiver are discussed. Such results are
based on measurements of their complex dielectric constant. We
evaluate that a 0.33 cm thick Teflon slab introduces negligible
spurious polarization ($< 2.6 \times 10^{-5}$ in transmission and
$< 6 \times 10^{-7}$ in emission), even minimizing the leakage ($<
10^{-8}$ from $Q$ to $U$ Stokes parameters, and viceversa) and the
depolarization ($\sim 1.3 \times 10^{-3}$).
\end{abstract}
\ocis{120.5410, 350.4010, 230.5440, 350.1260, 350.5500.}


\section{Introduction}
\label{intro}
The last decade has been characterized by the growing interest
into the Cosmic Microwave Background Polarization (CMBP), that has
stimulated the design of polarimeters featured by low systematic
effects and high sensitivity in microwave frequencies. The CMBP,
in fact, is among the most powerful tool to investigate the early
stages of the Universe \cite{review_CMBP}. Due to the faint
expected signal (a few $\mu$K over the $\sim$ 3 K unpolarized
component), any instrumental effect that can produce spurious
polarization must be analyzed in detail in order to minimize its
impact. Weak signals call for high sensitive radiometers that are
usually realized by cooling down the front-end to cryogenic
temperatures by means of cryostats. In Radioastronomy, homogeneous
and low loss dielectric slabs are used to allow the signal to
enter the cryostat (e.g. see experiments in Ref.
\citeonline{bar_sport,mario,polar,bicep,compass,dasi,masi}).
Usually, such materials are considered isotropic and then not
generating instrumental polarization. Hence, no analysis is
performed to study polarization effects related to isotropic flat
dielectrics. This work arises from the need to investigate the
linear instrumental polarization introduced by flat vacuum windows
in microwave polarimeters when considered isotropic, thus
providing the relationship with the material properties through
either the complex dielectric constant or the complex index of
refraction. However, the results reported in this paper are not
exhaustive as far as the complete design of vacuum windows is
concerned. Other effects may occur to enhance the systematics,
such as possible intrinsic or manufacturing induced birefringence
of the sample \cite{november}, and the present work should be
considered as a first attempting to investigate and minimize the
spurious instrumental polarization that may arise from flat
isotropic dielectric slabs. We find that a uniform diffuse
radiation does not generate instrumental polarization, which,
instead, is generated by anisotropic components (e.g. CMB). We
also estimate the amount of the effects in the specific case of
isotropic dielectric which works in Ka frequency band, centered at
32 GHz, an interesting band in microwave cosmology.
\\
The paper is organized as follows: in Section
(\ref{is_diel_sp_pol}) we present the theoretical model used to
derive the polarization effects introduced by flat isotropic
dielectrics, while in Section~(\ref{is_diel}) we present the total
intensity analysis which provides reflectance, transmittance and
absorptance of the materials. Measurements of their complex
dielectric constants are presented in section (\ref{micro_test})
and, finally, estimates of the spurious effects predicted by the
model for the dielectrics experimentally investigated are reported
in section (\ref{application}).
\section{Instrumental polarization by flat isotropic dielectric slab}
\label{is_diel_sp_pol}
\begin{figure} [h]
    \vspace{10 mm}
    \centerline{{\includegraphics[width=3in,height=3in]{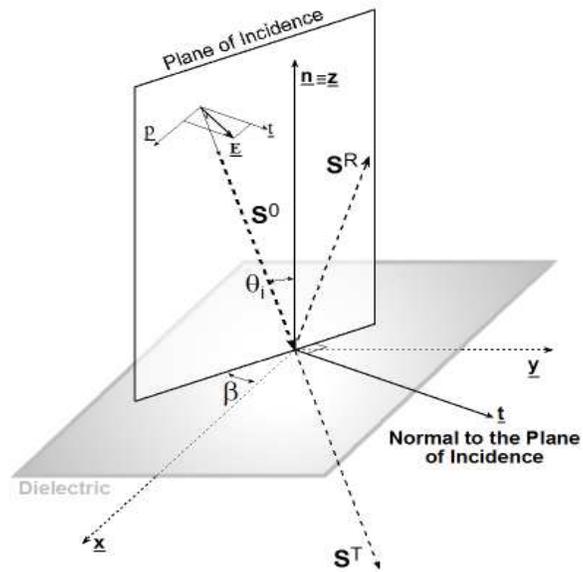}}}
    \caption{Reference frame: $\textbf{\underline{x}}$,
            $\textbf{\underline{y}}$ and $\textbf{\underline{z}}$ define the laboratory
            or feed-horn reference frame. $\beta$ is the angle between the plane of incidence
            and the $\textbf{\underline{x}}$ axis.}
    \label{refFrameFig}
\end{figure}

The instrumental polarization can be evaluated from the Stokes
parameter equations\cite{collett}:
\begin{eqnarray}\label{stokes_1}
    I^{x} & \propto &\langle |E_{p}^{x}|^{2} \rangle + \langle |E_{t}^{x}|^{2} \rangle
    \\ \label{stokes_1_1}
    Q^{x} & \propto &\langle |E_{p}^{x}|^{2} \rangle - \langle |E_{t}^{x}|^{2} \rangle
    \\ \label{stokes_1_2}
    U^{x} & \propto &\langle 2\Re\{E_{p}^{x}E_{t}^{x^{*}}\} \rangle \\ \label{stokes_1_3}
    V^{x} & \propto &\langle 2\Im\{E_{p}^{x}E_{t}^{x^{*}}\} \rangle
\end{eqnarray}
where $E_{p}^{x}$ and $E_{t}^{x}$ are the complex parallel and
perpendicular components of the electric field with respect to the
incident plane, $\langle \rangle$ denotes the time average
(henceforth we will omit it for clarity),
 and $x = R,T$ indicates the reflection and transmission terms,
respectively.
\\
The transmission case writes
\begin{equation}
    E_{j}^{T} = T_{j} E_{j} \label{stokes_9}
\end{equation}
where $j=p,t$ denotes the parallel and perpendicular components of
the incoming $E_{j}$ electric field and $T_{j}$ is its complex
transmission coefficient due to the flat dielectric slab
\cite{born,brehat,piro,azzam}. Using Eq. (\ref{stokes_9}) in
Eq.~(\ref{stokes_1})--(\ref{stokes_1_3}), the Stokes parameters of
the transmitted wave, computed in the reference frame defined by
the plane of incidence, are given by
\begin{equation}\label{stokes_12}
    {\bf{S}}^{T} = {\bf M}^{T} {\bf S}^{0}
\end{equation}
where $\textbf{S}^{0} \equiv [I^{0}, Q^{0}, U^{0}, V^{0}]$ and
$\textbf{S}^{T} \equiv [I^{T}, Q^{T}, U^{T}, V^{T}]$ are the
Stokes parameters of the incoming and transmitted radiation
respectively, and
\begin{equation} \label{stokes_13}
\mathbf{M}^{T} = \left(
\begin{array}{cccc}
    \frac{1}{2}(|T_{p}|^{2}+|T_{t}|^{2}) & \frac{1}{2}(|T_{p}|^{2}-|T_{t}|^{2}) & 0 & 0 \\
    \frac{1}{2}(|T_{p}|^{2}-|T_{t}|^{2}) & \frac{1}{2}(|T_{p}|^{2}+|T_{t}|^{2}) & 0 & 0 \\
    0 & 0 & \Re\{T_{p} T_{t}^{*}\} & \Im\{T_{p} T_{t}^{*}\}\\
    0 & 0 & - \Im\{T_{p} T_{t}^{*}\} & \Re\{T_{p} T_{t}^{*}\}\\
\end{array} \right)
\end{equation}
is the Mueller matrix \cite{collett}. Since generally $T_{p} \neq
T_{t}$ and $T_{p}T_{t}^{*} \neq 0$, a flat isotropic dielectric
slab generates cross-contamination between $I$ and $Q$, and $U$
and $V$ parameters.
\\
An interesting case is that of unpolarized incoming radiation,
defined by $I_{\rm un}^{0}$. In this case, the outgoing radiation
is featured by
\begin{equation} \label{stokes_16}
\mathbf{S}^{T} = \left(
\begin{array}{c}
    \frac{1}{2}(|T_{p}|^{2} + |T_{t}|^{2}) I_{\rm un}^{0}     \\
    \frac{1}{2}(|T_{p}|^{2} - |T_{t}|^{2}) I_{\rm un}^{0}     \\
    0                                           \\
    0                                           \\
\end{array} \right)
\end{equation}
Therefore, an instrumental polarization is generated and the
contamination affects only $Q$. Hereafter, we call it {\it
spurious polarization}.
\\
Similar considerations hold for the reflection case, by replacing
$T_{j}$ by $R_{j}$.
\\
It is worth noting that $\mathbf{M}^{T}$ (and $\mathbf{M}^{R}$) is
the product of the Mueller matrices of polarizer and retarder
\cite{collett}, allowing the description of a flat isotropic
dielectric slab as a combination of these two polarizing devices.
\\
About the effects generated by the emitted radiation, we adopt an
approach based on the radiation power rather than on the electric
field. The signal emitted by the slab is thermal noise
characterized by a continuum spectrum related to the physical
temperature of the dielectric ${\rm T\!}_{\rm ph}$. In the
microwave frequency domain, by adopting the Rayleigh-Jeans
approximation, the brightness temperature of a thermal source is
proportional to its physical temperature \cite{kraus}. Thus, the
two intensity components (parallel and perpendicular) of the
emitted signal can be computed as
\begin{equation}
   |E_j^{\varepsilon}|^2 \propto \varepsilon\!_j {\rm T\!}_{\rm ph}
\end{equation}
where $\varepsilon\!_{j}$ is the Emittance for the $j$-component.
Assuming that the slab is in thermal equilibrium, the equivalence
between Emittance $\varepsilon\!_{j}$ and Absorptance $A_{j}$
holds\cite{jordan}
\begin{equation} \label{stokes_22}
    \varepsilon \!_{j}(\nu,\theta_{i}) \equiv A_{j}(\nu,\theta_{i}) = 1 -
    |R_{j}(\nu,\theta_{i})|^{2} - |T_{j}(\nu,\theta_{i})|^{2}
\end{equation}
allowing us the computation of the emission coefficients from
those of reflection and transmission.
Due to the thermal noise nature of the emitted components, these
can be considered uncorrelated\cite{jordan}, so that the Stokes
parameters write
\begin{equation} \label{stokes_25}
\mathbf{S^{\varepsilon}} = \left(
\begin{array}{c}
    \frac{1}{2}(\varepsilon_{p}+\varepsilon_{t}) {\rm T\!}_{\rm ph}    \\
    \frac{1}{2}(\varepsilon_{p}-\varepsilon_{t}) {\rm T\!}_{\rm ph}   \\
    0                                \\
    0                                \\
\end{array} \right)
\end{equation}
giving the interesting result that the thermal noise injected by
the flat slab is partially polarized. Once again, the
contamination affects only $Q$.
\\
In case of unpolarized incoming radiation, it is convenient to
describe the instrumental polarization through the equations
\begin{eqnarray}
    Q^{R}_{S\!P}           & = & S\!P^{R}\, I_{\rm un}^{0} \label{stokes_26_II}    \\
    Q^{T}_{S\!P}           & = & S\!P^{T}\, I_{\rm un}^{0} \label{stokes_26_bis_II} \\
    Q^{\varepsilon}_{S\!P} & = & S\!P^{\varepsilon}\, {\rm T\!}_{\rm ph} \label{stokes_26_ter_II}
\end{eqnarray}
with the {\it spurious polarization coefficients} $S\!P$ given by
\begin{eqnarray}\label{stokes_26}
    S\!P^{R} & = & \frac{1}{2} (|R_{p}|^{2} - |R_{t}|^{2}) \\\label{stokes_26_bis}
    S\!P^{T} & = & \frac{1}{2} (|T_{p}|^{2} - |T_{t}|^{2}) \\\label{stokes_26_ter}
    S\!P^{\varepsilon} & = & \frac{1}{2} (\varepsilon_{p} - \varepsilon_{t})
\end{eqnarray}
While $|R_j|^{2}$ and $|T_j|^{2}$ can be measured, it is generally
hard to measure $\varepsilon \!_j$ and, in turn, the coefficient
$S\!P^{\varepsilon}$. However, it is straightforward that the
relation
\begin{equation}\label{somme_spurie}
    S\!P^{R} + S\!P^{T} + S\!P^{\varepsilon} = 0
\end{equation}
holds, providing thus a way to evaluate $S\!P^{\varepsilon}$.
\\
Interesting special case is that of perpendicular incident
radiation ($\theta_i = 0$ in Fig. \ref{refFrameFig}), for which
$|R_p|^{2} = |R_t|^{2}$ and $|T_p|^{2} = |T_t|^{2}$, then
\begin{equation}
    S\!P^{x}(\theta_{i} = 0) = 0 \label{sp_0_90}
\end{equation}
that is no spurious polarization is generated.
\\
Frequently, the dielectric slab is placed in front of a collecting
system (e.g. a feed-horn which we assume aligned with the slab
axis), thus it is important to evaluate the instrumental
polarization propagating inside the receiver. Contributions come
from the radiation transmitted and emitted by the dielectric. In
fact, a further signal can be generated by reflection due to the
signal emitted by the receiver toward the free space and
backscattered by the slab. However, this occurs at angles of
incidence $\theta_{i} \sim 0$, thus implying
negligible~$S\!P^{R}$.
\\
The Eq. (\ref{stokes_12}) and~(\ref{stokes_25}) provide the
instrumental polarization due to the radiation coming from a given
direction and in the reference frame defined by the plane of
incidence and its normal (see Fig.~\ref{refFrameFig}).
\\
Such signals must be evaluated in the laboratory reference frame
(e.g. the Antenna Reference Frame: $A\!R\!F$), by accounting for
the rotation of the azimuthal angle $\beta$. By referring to
Fig.~\ref{refFrameFig}, if $\textbf{S}_{\rm ARF}^{0}$ defines the
Stokes parameters of the incoming radiation in the $A\!R\!F$, the
polarization state of the transmitted component can be described
as \cite{collett}:
\begin{equation} \label{s_t_arf}
    \textbf{S}_{\rm ARF}^{T}(2 \beta) =
    \textbf{R}(-2 \beta) \mathbf{M}^{T} \textbf{R}(2 \beta) \; \textbf{S}_{\rm ARF}^{0} =
    \textbf{H}^{T}(2 \beta) \; \textbf{S}_{\rm ARF}^{0}
\end{equation}
where the $\textbf{H}^{T}(2 \beta)$ matrix is:
\begin{equation}
\label{m_t_2beta_matrix} \left(
\begin{array}{cccc}
 T                     & S\!P^{T}\cos(2 \beta) & S\!P^{T}\sin(2 \beta) & 0 \\
 S\!P^{T}\cos(2 \beta) & T\cos^{2}(2 \beta) + M_{33}^{T}\sin^{2}(2 \beta) & (T-M_{33}^{T})\sin(2 \beta)\cos(2 \beta) & - L^{T}\sin(2 \beta)  \\
 S\!P^{T}\sin(2 \beta) & (T-M_{33}^{T})\sin(2 \beta)\cos(2 \beta) & T\sin^{2}(2 \beta) + M_{33}^{T}\cos^{2}(2 \beta) & L^{T}\cos(2 \beta)  \\
 0                     & L^{T}\sin(2 \beta)     & - L^{T}\cos(2 \beta) & M_{33}^{T}
\end{array} \right)
\end{equation}
and
\begin{eqnarray}
    T           & = & \frac{1}{2} ( |T_{p}|^{2} + |T_{t}|^{2}) \label{transmitt}\\
    L^{T}       & = & \Im\{T_{p} T_{t}^{*}\} \label{leakage_U}\\
    M_{33}^{T}  & = & \Re\{T_{p} T_{t}^{*}\}  \label{m33_stokes}
\end{eqnarray}
where $\textbf{R}(2 \beta)$ is the Mueller matrix for rotation
\cite{collett} and $T$ is the Transmittance of the slab. Given an
input Stokes parameter Y$^{0}$, we define its contamination over
the X output as X$_{\rm Y^{0}}$: such a cross-term is the
$\textit{leakage}$. Thus, by definition, the transmission function
of the X$^{0}$ parameter is X$_{\rm X^{0}}$.
\\
About the emitted component, in the $A\!R\!F$ the signal expected
is:
\begin{equation} \label{m_epsilon_2beta_matrix}
\mathbf{S}_{\rm ARF}^{\varepsilon}(2 \beta) = \left(
\begin{array}{c}
    \varepsilon {\rm T\!}_{\rm ph}    \\
    S\!P^{\varepsilon}\cos(2 \beta) {\rm T\!}_{\rm ph}   \\
    S\!P^{\varepsilon}\sin(2 \beta) {\rm T\!}_{\rm ph}   \\
    0                                \\
\end{array} \right)
\end{equation}
where $\varepsilon=\frac{1}{2}(\varepsilon_{p}+\varepsilon_{t})$
is the Emittance of the slab.
\\
It is worth noting that each quantity hitherto discussed is
well-defined for values of the polar angle up to $\frac{\pi}{2}$
rad.
\\
The effect on the spurious polarization due to the incoming
unpolarized diffuse signal characterized by a brightness
temperature distribution \cite{kraus} ${\rm T\!}_{\rm b}(\nu,
\theta, \beta)$ can be computed by integrating the spurious
components ($S\!P^{T}$) of Eq. (\ref{s_t_arf}) all over the
directions. Such integrations lead to the following outputs for
$Q$ and $U$ in antenna temperature \cite{kraus}:
\begin{eqnarray}
    Q^{T}_{S\!P}(d) & = & \frac{1}{\Omega_{A}^{\rm FF}\Delta\nu} \int_{\Delta\nu}
    \bigg [ \int_{0}^{\pi/2}\textrm{d}\theta S\!P^{T}(d,\nu,\theta^{\;\prime})
                P_{n}^{\rm FF}(\theta) \sin(\theta) \int_{0}^{2 \pi}\textrm{d}\beta
                {\rm T\!}_{\rm b}(\nu,\theta,\beta) \cos(2\beta) \bigg]
                \textrm{d}\nu \nonumber  \\
                \label{stokes_27} \\
    U^{T}_{S\!P}(d) & = & \frac{1}{\Omega_{A}^{\rm FF}\Delta\nu} \int_{\Delta\nu}
    \bigg [ \int_{0}^{\pi/2}\textrm{d}\theta S\!P^{T}(d,\nu,\theta^{\;\prime})
                P_{n}^{\rm FF}(\theta) \sin(\theta) \int_{0}^{2 \pi}\textrm{d}\beta {\rm T\!}_{\rm b}(\nu,\theta,\beta)
                \sin(2\beta) \bigg] \textrm{d}\nu   \nonumber
                \label{stokes_27_1} \\
\end{eqnarray}
where $\Delta \nu$ is the frequency bandwidth, $P_{n}^{{\rm FF}}$
and $\Omega_{A}^{{\rm FF}}$ are the normalized co-polar pattern
and the antenna solid angle in far field regime \cite{kraus}
respectively; $\theta^{\; \prime} (\theta)$ is the angle of
incidence on the slab of the radiation coming from the (far field)
direction $\theta$: if the slab is in the near field of the
antenna, then $\theta^{\; \prime} < \theta$ (flat slab and antenna
are assumed coaxial, so $\theta_i \equiv \theta$), otherwise in
far field $\theta^{\; \prime} \equiv \theta$. In order to
emphasize the effect introduced by the slab, in Eq.
(\ref{stokes_27}) and (\ref{stokes_27_1}) we assume an ideal
feed-horn featured by a $\beta$-symmetric co-polar pattern (i.e.
null cross-polarization), the feed horn spectral transfer function
constant all over the frequency band ($P_{n}(\nu,\theta,\beta)
\equiv P_{n}(\theta) $) and negligible edge effects between the
slab and the feed aperture in the case of near field position.
\\
Writing the brightness temperature with respect to its mean and
anisotropy components
\begin{equation}\label{stokes_29}
    {\rm T\!}_{\rm b}(\nu,\theta,\beta) = {\rm T\!}_{\rm b,0}(\nu) + \Delta \! {\rm T\!}_{\rm b}(\nu,\theta,\beta)
\end{equation}
makes null the contribution of the mean term in the Eq.
(\ref{stokes_27})--(\ref{stokes_27_1}), which thus become
\begin{eqnarray}
    Q^{T}_{S\!P}(d) & = & \frac{1}{\Omega_{A}^{\rm FF}\Delta\nu} \int_{\Delta \nu}
                     \textrm{d}\nu \int_{0}^{\pi/2}\textrm{d}\theta
                     S\!P^{T}(\theta^{\;\prime}) P_{n}^{\rm FF}(\theta) \sin(\theta)
                     \int_{0}^{2 \pi}\textrm{d}\beta \Delta \! {\rm T\!}_{\rm b}(\nu,\theta,\beta)
                     \cos(2\beta) \nonumber \\
                     \label{stokes_30} \\
    U^{T}_{S\!P}(d) & = & \frac{1}{\Omega_{A}^{\rm FF}\Delta\nu} \int_{\Delta \nu}
                     \textrm{d}\nu \int_{0}^{\pi/2}\textrm{d}\theta
                     S\!P^{T}(\theta^{\;\prime}) P_{n}^{\rm FF}(\theta)
                     \sin(\theta) \int_{0}^{2 \pi}\textrm{d}\beta
                     \Delta \! {\rm T\!}_{\rm b}(\nu,\theta,\beta) \sin(2\beta)  \nonumber \\
                     \label{stokes_30_bis}
\end{eqnarray}
This gives the important result that a flat isotropic dielectric
slab can generate spurious polarization only in presence of
anisotropic incoming signal. Similarly, in axisymmetric optics
\cite{spurious,carretti_2} the instrumental polarization is
generated only by incoming anisotropic radiation. This is
particularly relevant for the CMB, whose anisotropy is low
($\Delta \! {\rm T\!}_{\rm b}/ {\rm T\!}_{\rm b} \sim~10^{-5}$)
\cite{map,boomerang}.
\\
The spurious polarization due to the emission, can be computed by
replacing ${\rm T\!}_{\rm b}$ with the physical temperature and,
from Eq. (\ref{m_epsilon_2beta_matrix}), the correlated component
of the thermal noise collected by the feed is
\begin{eqnarray}
    Q^{\varepsilon}_{S\!P}(d) = \frac{1}{\Omega_{A}^{\rm N\!F}\Delta\nu} \int_{\Delta \nu}
                      \textrm{d}\nu \int_{0}^{\pi/2}\textrm{d}\theta
                    S\!P^{\varepsilon}(d,\nu,\theta) P_{n}^{\rm N\!F}(\theta)
                    \sin(\theta) \int_{0}^{2 \pi}\textrm{d}\beta
                    \; \Delta \! {\rm T\!}_{\rm ph}(\theta,\beta) \cos(2\beta) \nonumber \\ \label{stokes_36}\\
    U^{\varepsilon}_{S\!P}(d) = \frac{1}{\Omega_{A}^{\rm N\!F}\Delta\nu} \int_{\Delta \nu}
                    \textrm{d}\nu \int_{0}^{\pi/2}\textrm{d}\theta
                    S\!P^{\varepsilon}(d,\nu,\theta) P_{n}^{\rm N\!F}(\theta)
                    \sin(\theta) \int_{0}^{2 \pi}\textrm{d}\beta
                    \; \Delta \! {\rm T\!}_{\rm ph}(\theta,\beta) \sin(2\beta) \nonumber \\ \label{stokes_36_bis}
\end{eqnarray}
where $P^{\rm N\!F}_n$ is the normalized co-polar pattern in near
field at the slab position if it is placed in the near field of
the antenna. Even in this case, the contribution of the mean
component is null and the spurious polarization equations are like
for Eq. (\ref{stokes_30})--(\ref{stokes_30_bis}), but with the
anisotropic thermal component $\Delta \! {\rm T\!}_{\rm ph}$
instead of $\Delta \! {\rm T\!}_{\rm b}$, and $P^{\rm F\!F}_n$
replaced by $P^{\rm N\!F}_n$.
\\
Since both the transmitted and emitted signals are not correlated,
due to the additive property of the Stokes parameters
\cite{collett} the total spurious polarizations is
\begin{eqnarray}\label{tot_sp_1}
    Q_{S\!P} & = & Q^{T}_{S\!P} + Q^{\varepsilon}_{S\!P} \\
    U_{S\!P} & = & U^{T}_{S\!P} + U^{\varepsilon}_{S\!P}
\end{eqnarray}
From the matrix $\textbf{H}^{T}(2 \beta)$, the depolarization of
$Q$ and $U$ can be estimated as:
\begin{eqnarray}
    D_{Q^{T}}     & = & \frac{Q^{0} - Q^{T}}{Q^{0}} = (1 - T) \cos^{2}(2\beta) + (1 - M_{33}^{T}) \sin^{2}(2\beta) \label{depol_Q} \\
    D_{U^{T}}     & = & \frac{U^{0} - U^{T}}{U^{0}} = (1 - T) \sin^{2}(2\beta) + (1 - M_{33}^{T}) \cos^{2}(2\beta) \label{depol_Q}
\end{eqnarray}
The loss of the Q-signal, $D_{Q^{T}}$, could be evaluated in
percentage term with respect to the measured $Q^{0}_{m}$ in
absence of the slab. The result is:
\begin{equation}\label{eta_q_t}
    \eta_{Q^{T}}(d) = \frac{\frac{1}{\Omega_{A}^{\rm F\!F} \Delta\nu} \int_{\Delta \nu}
                    \rm{d}\nu \int_{0}^{\pi/2} \rm{d}\theta \; P_{n}^{F\!F}(\theta)
                    \sin(\theta) \int_{0}^{2 \pi} \rm{d}\beta \; D_{Q^{T}}(d,\nu,\theta^{\; \prime},\beta)
                    Q^{0}(\nu,\theta,\beta)}{\frac{1}{\Omega_{A}^{\rm F\!F} \Delta\nu}
                    \int_{\Delta \nu} \rm{d}\nu \int_{0}^{\pi/2} \rm{d}\theta \; P_{n}^{F\!F}(\theta)
                    \sin(\theta) \; \int_{0}^{2 \pi} \rm{d}\beta Q^{0}(\nu,\theta,\beta)}
\end{equation}
The same for $U$. Assuming a constant incoming signal
($Q^{0}(\nu,\theta,\beta) = Q^{0}$), a zero-order estimate of Eq.
(\ref{eta_q_t}) is given by:
\begin{eqnarray}
    \eta_{Q^{T}}(d) & = & \frac{2\pi}{\Omega_{A}^{\rm F\!F}} \int_{0}^{\pi/2} \tilde{D}_{Q^{T}}(d,\theta^{\; \prime})
                   P_{n}^{F\!F}(\theta) \sin(\theta) \; \rm{d}\theta \label{eta_q_t_q0_const} \\
    \tilde{D}_{Q^{T}}(d,\theta^{\; \prime}) & = & \frac{1}{2 \pi} \int_{0}^{2\pi} \langle D_{Q^{T}}(d,\nu,\theta^{\;
                    \prime},\beta) \rangle_{\nu} \; \rm{d}\beta = \nonumber \\
                    & = & \frac{1}{2} \left [ 1 - \langle T(d,\nu,\theta^{\; \prime}) \rangle_{\nu}
                    + 1 - \langle M_{33}^{T}(d,\nu,\theta^{\; \prime}) \rangle_{\nu}\right ] \label{d_tilde}
\end{eqnarray}
where $\langle\rangle_{\nu}$ is the in-band average.
\\
The $Q_{U^{0}}^{T}$ leakage term results:
{\setlength\arraycolsep{0pt}
\begin{eqnarray} \label{q_u0}
Q_{U^{0}}^{T}(d)
                & \; = \; & \frac{1}{\Omega_{A}^{\rm FF}\Delta\nu} \int_{\Delta \nu} \rm{d} \nu
                            \int_{0}^{\pi/2}\rm{d}\theta \; \bigg [T(d,\nu,\theta^{\; \prime}) -
                            M_{33}^{T}(d,\nu,\theta^{\; \prime}) \bigg] P_{n}^{\rm FF}(\theta)
                            \sin(\theta) \nonumber\\
                & \times &  \int_{0}^{2 \pi} \rm{d}\beta \; \sin(2 \beta) \cos(2 \beta)
                            \Delta \! U^{0}(\nu,\theta,\beta)
\end{eqnarray}}
The same for $U_{Q^{0}}^{T}$ by replacing $Q$ with $U$. Note that
a non null result arises if the incoming signal is anisotropic.
\\
Finally, the leakage due to $V^{0}$ is given by:
\begin{eqnarray}
        Q_{V^{0}}^{T}(d)
               & = & - \frac{1}{\Omega_{A}^{\rm FF}\Delta\nu} \int_{\Delta \nu} \rm{d} \nu
                            \int_{0}^{\pi/2}\rm{d}\theta L^{T}(d,\nu,\theta^{\; \prime})
                            P_{n}^{\rm FF}(\theta) \sin(\theta) \int_{0}^{2 \pi} \rm{d}\beta \;
                            \sin(2 \beta) \Delta \! V^{0}(\nu,\theta,\beta) \nonumber\\ \label{q_v0}\\
        U_{V^{0}}^{T}(d)
               & = & \frac{1}{\Omega_{A}^{\rm FF}\Delta\nu} \int_{\Delta \nu} \rm{d} \nu
                            \int_{0}^{\pi/2}\rm{d}\theta L^{T}(d,\nu,\theta^{\; \prime})
                            P_{n}^{\rm FF}(\theta) \sin(\theta) \int_{0}^{2 \pi} \rm{d}\beta \;
                            \cos(2 \beta) \Delta \! V^{0}(\nu,\theta,\beta)  \nonumber\\ \label{u_v0}
\end{eqnarray}
Once again, a leakage is generated only in case of anisotropic
incoming signal.
\section{Total intensity analysis}
\label{is_diel}
The effects on the total intensity signal can be evaluated by
estimating both transmission and reflection properties of the slab
(see for details Ref. \citeonline{born,brehat,piro,azzam}).
\\
The incoming signal collected by the system slab-feed is given by
(in antenna temperature):
\begin{equation}\label{trans_out}
    {\rm T}_{A}^{T}(d) = \frac{1}{\Omega_{A}^{{\rm FF}} \Delta \nu} \int_{0}^{\pi/2} \textrm{d}\theta
                \int_{0}^{2\pi} \textrm{d}\beta \int_{\Delta \nu} \textrm{d}\nu \; T(d,\nu,\theta^{\; \prime})
                {\rm T\!}_{\rm b}(\nu,\theta,\beta) \; P_{n}^{{\rm FF}}(\theta)
                \sin(\theta)
\end{equation}
where $T$ is the Transmittance of the slab.
\\
An estimate of the effect is given by the relative transmitted
signal $\Lambda(d) = {\rm T}_{A}^{T}(d) / {\rm T\!}_{0}$ in the
simple case of an isotropic input signal ${\rm T\!}_{\rm
b}(\nu,\theta,\beta) = {\rm T\!}_{0}$:
\begin{equation}\label{trans_out_e}
    \Lambda(d) = \frac{2 \pi}{\Omega_{A}^{\rm FF}} \int_{0}^{\pi/2}
                 \langle T(d,\nu,\theta^{\; \prime}) \rangle_\nu
                P_{n}^{\rm FF}(\theta) \sin(\theta) \textrm{d}\theta
\end{equation}
If the feed horn directivity is high, most of the signal is
collected close to $0^{\circ}$. In the limit of low loss
dielectrics, the maxima of $\Lambda(d)$ will be identified by
integer multiples of the well known thickness $d = d_{0}
\equiv\lambda/2n$, which identifies the Transmittance maxima for
null incidence \cite{born}, where $n$ is the real part of the
complex index of refraction.
\\
Similarly, the thermal noise injected by the dielectric can be
computed when it is in thermal equilibrium at the physical
temperature ${\rm T\!}_{\rm ph}$. Its emission is that of a
greybody at temperature ${\rm T\!}_{\rm ph}$ featured by an
Emittance $\varepsilon(\theta_{i}) \equiv A(\theta_{i}) =
1-[R(\theta_{i})+T(\theta_{i})]$, where $R = \frac{1}{2} (
|R_{p}|^{2} + |R_{t}|^{2})$ is the Reflectance of the slab. Since
here we consider the microwave frequency domain, the
Rayleigh-Jeans approximation can be adopted. Hence, in term of
brightness temperature \cite{kraus}, the thermal noise emitted by
the slab is simply $\varepsilon(\theta_{i}) {\rm T\!}_{\rm ph}$.
Thus, the signal collected by the antenna is:
\begin{equation}\label{emiss_out}
    {\rm T}_{A}^{\varepsilon}(d) = \frac{1}{\Omega_{A}^{{\rm NF}} \Delta \nu} \int_{0}^{\pi/2} \textrm{d}\theta
               \int_{0}^{2\pi} \textrm{d}\beta \int_{\Delta \nu} \textrm{d}\nu \; \varepsilon (d,\nu,\theta) \;
               {\rm T\!}_{\rm ph}(\theta,\beta) P_{n}^{{\rm NF}}(\theta) \sin(\theta)
\end{equation}
where $P_{n}^{{\rm NF}}(\theta)$ is the Near Field normalized
co-polar pattern of the feed if the slab is placed at its near
field, and a thermally inhomogeneous but thermally stabilized slab
has been considered.
\\
In a conservative approach, the antenna noise temperature can be
estimated as:
\begin{equation}\label{emiss_out_sovrastima}
    {\rm T}_{A}^{\varepsilon}(d) \leq
                \langle \varepsilon(d,\nu,\theta) \rangle_{\nu}|_{{\rm MAX}(\theta)}
                \times {\rm T\!}_{\rm ph}(\theta,\beta)|_{{\rm MAX}(\theta,\beta)}
\end{equation}
where MAX($y$) stands for the maximum value over the quantity $y$.
\section{Microwave Tests}
\label{micro_test}
\begin{figure} [h]
    \includegraphics[width=2.5in,height=1.25in]{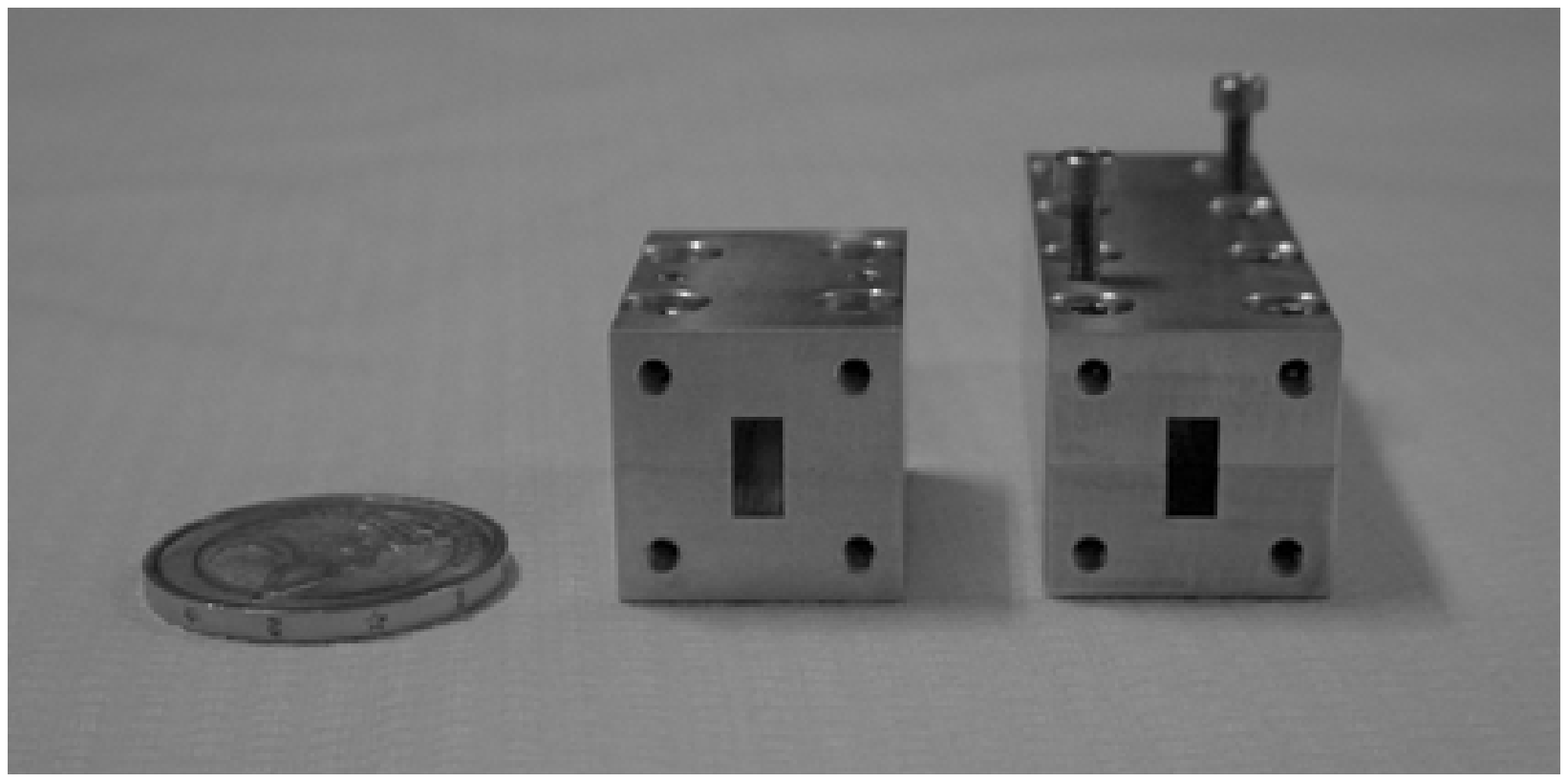} \hspace{30 mm}
    \includegraphics[width=1.25in,height=1.25in]{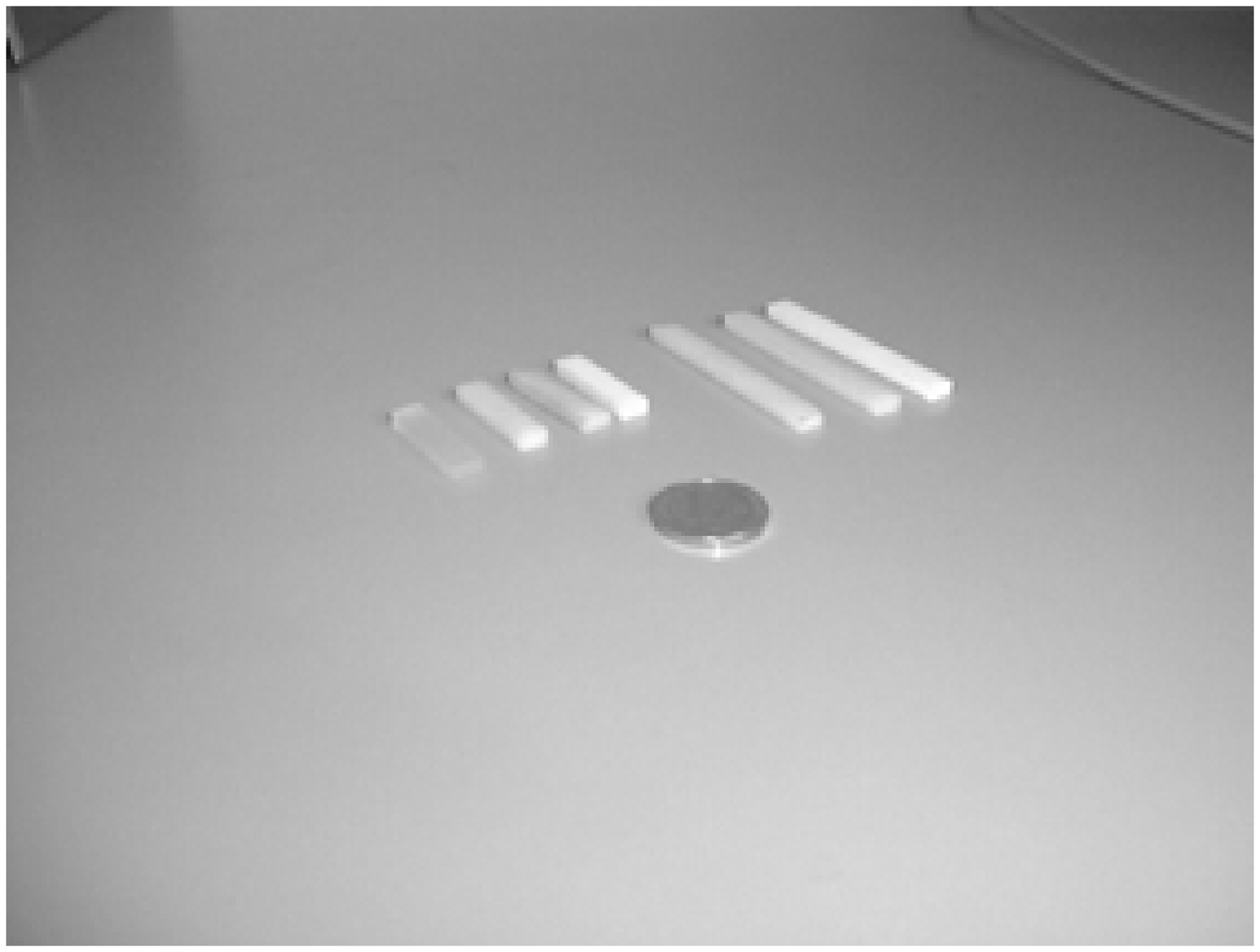}
    \caption{Waveguides and dielectric samples.}
    \label{misure_dielettrici_foto}
\end{figure}

\begin{figure} [h]
    \includegraphics[width=4in,height=3in]{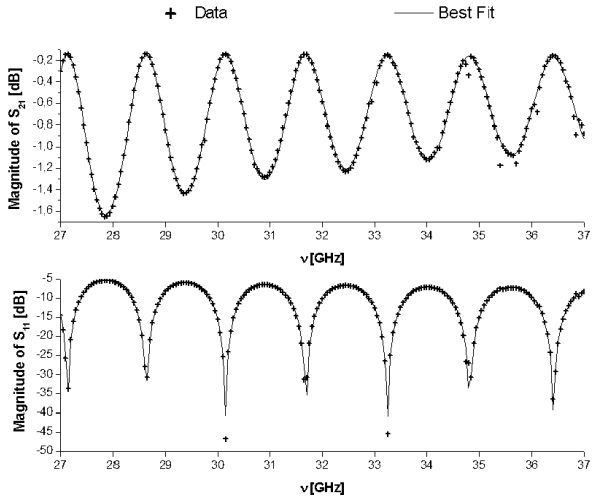}
    \caption{Measurements (+) and best fit (---) of scattering
            parameters of the Teflon 1-cut loaded waveguide.}
    \label{simulation_data}
\end{figure}

\begin{figure} [h]
    \centerline{\includegraphics[width=4in,height=1.5in]{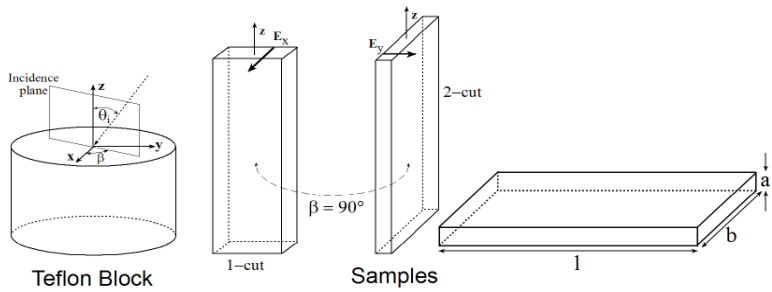}}
    \caption{Dielectric cuts. $\textbf{E}_x$ and $\textbf{E}_y$ are the electric field components vibrating
            along the $\textbf{x}$ and $\textbf{y}$ directions in the Antenna Reference
            Frame. $a$, $b$ and $l$ denote the sample sizes. 1--cut and 2--cut refers only to
            Teflon. See text for details.}
    \label{materials_size_fig}
\end{figure}

\begin{figure} [h]
    \includegraphics[width=2.8in,height=2.3in]{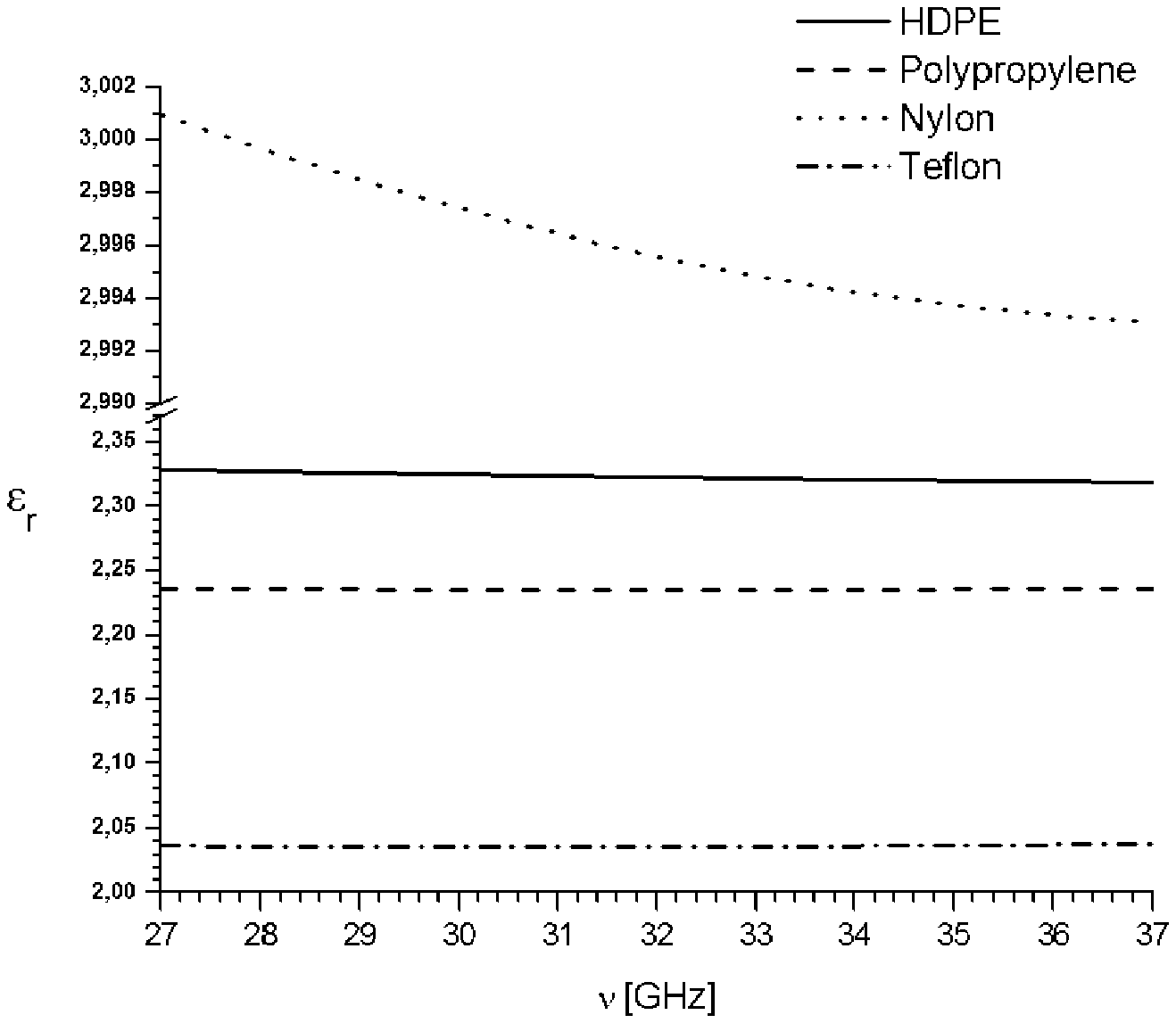} \hspace{15 mm}
    \includegraphics[width=2.7in,height=2.2in]{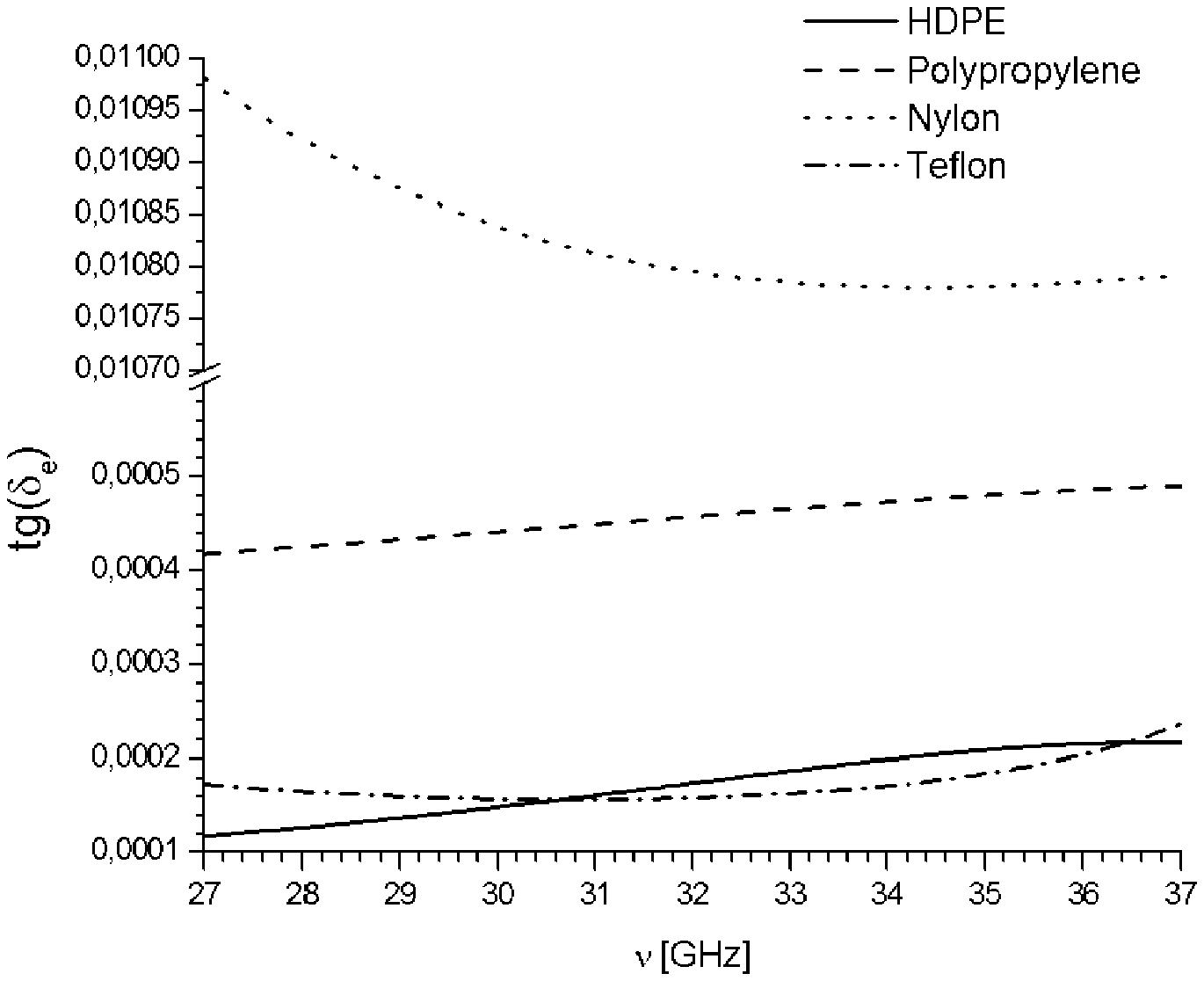}
        \\\\\\\\\\
    \includegraphics[width=2.7in,height=2.2in]{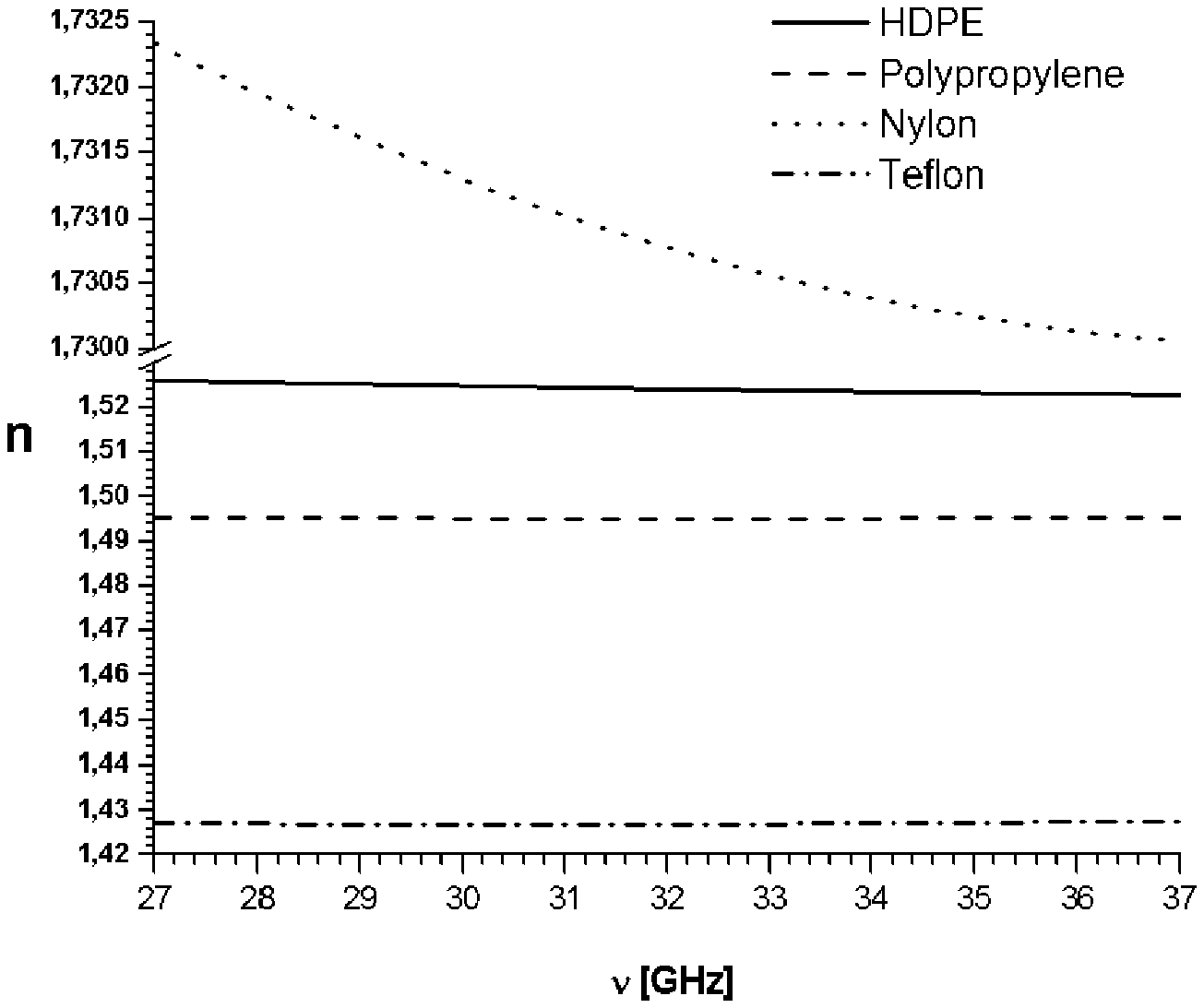} \hspace{15 mm}
    \includegraphics[width=2.7in,height=2.2in]{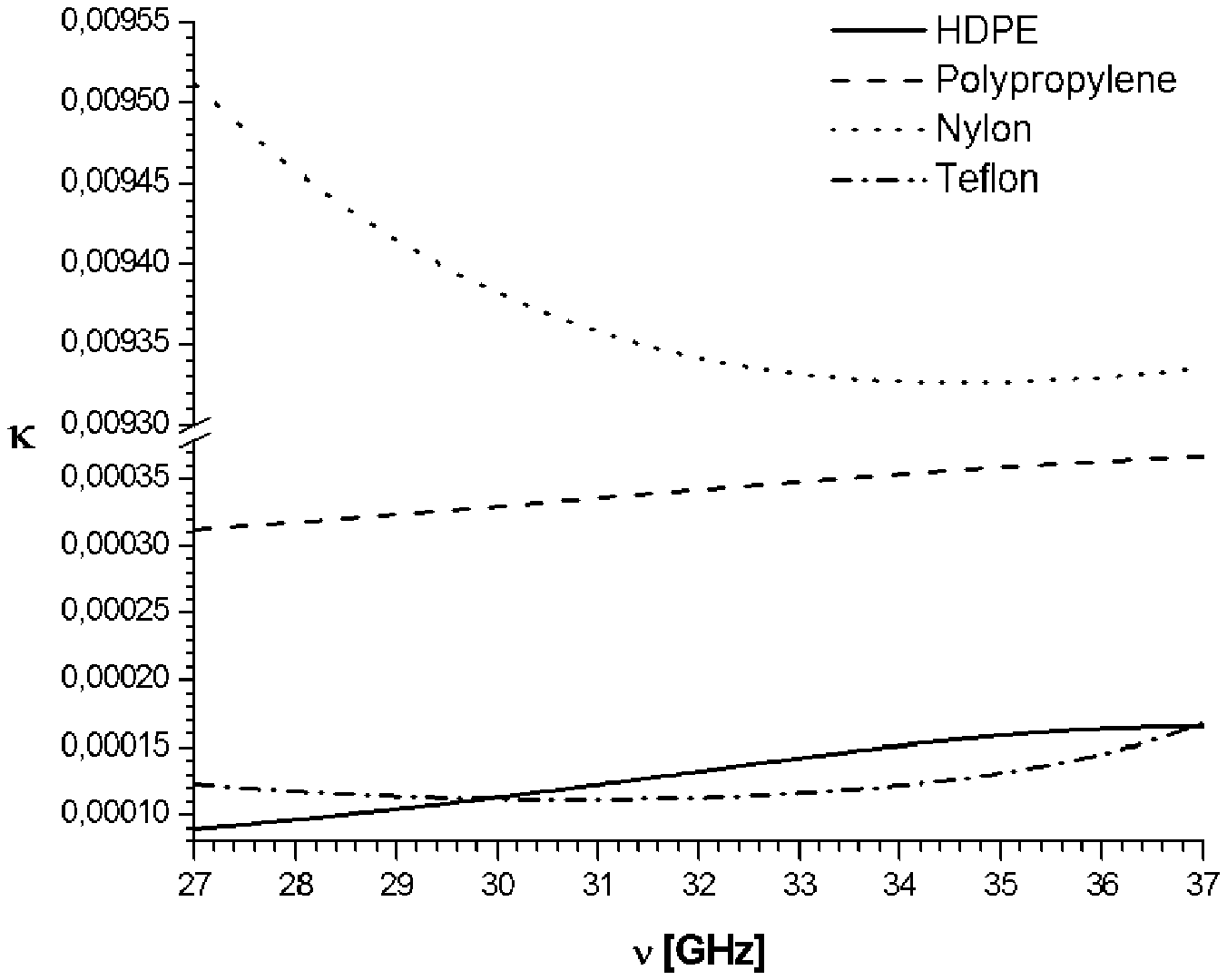}
    \caption{Complex dielectric constant and index of refraction of: HDPE, Polypropylene, Nylon and 1--cut
            Teflon. The measurements have been performed at T$_{\rm{ph}}$ = 300 K with
            $\sim$ 1$\%$ of precision for $\epsilon_r$ and $n$, and $\sim$ 30$\%$ for
            $\tan(\delta_e)$ and $\kappa$ (only for Nylon is 1$\%$).}
    \label{misure_dielettrici_1}
\end{figure}

\begin{figure} [h]
    \includegraphics[width=2.7in,height=2.2in]{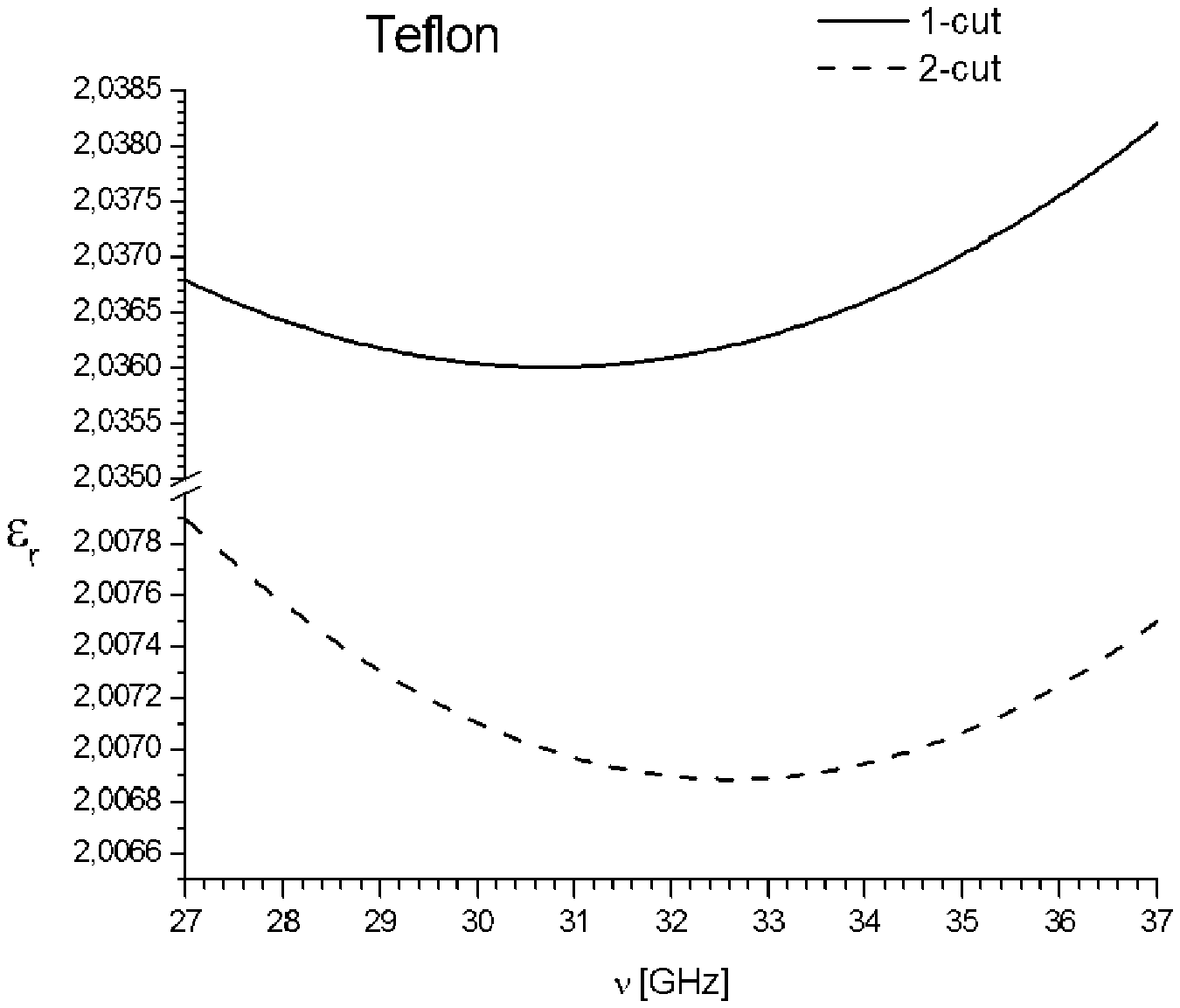} \hspace{15 mm}
    \includegraphics[width=2.7in,height=2.2in]{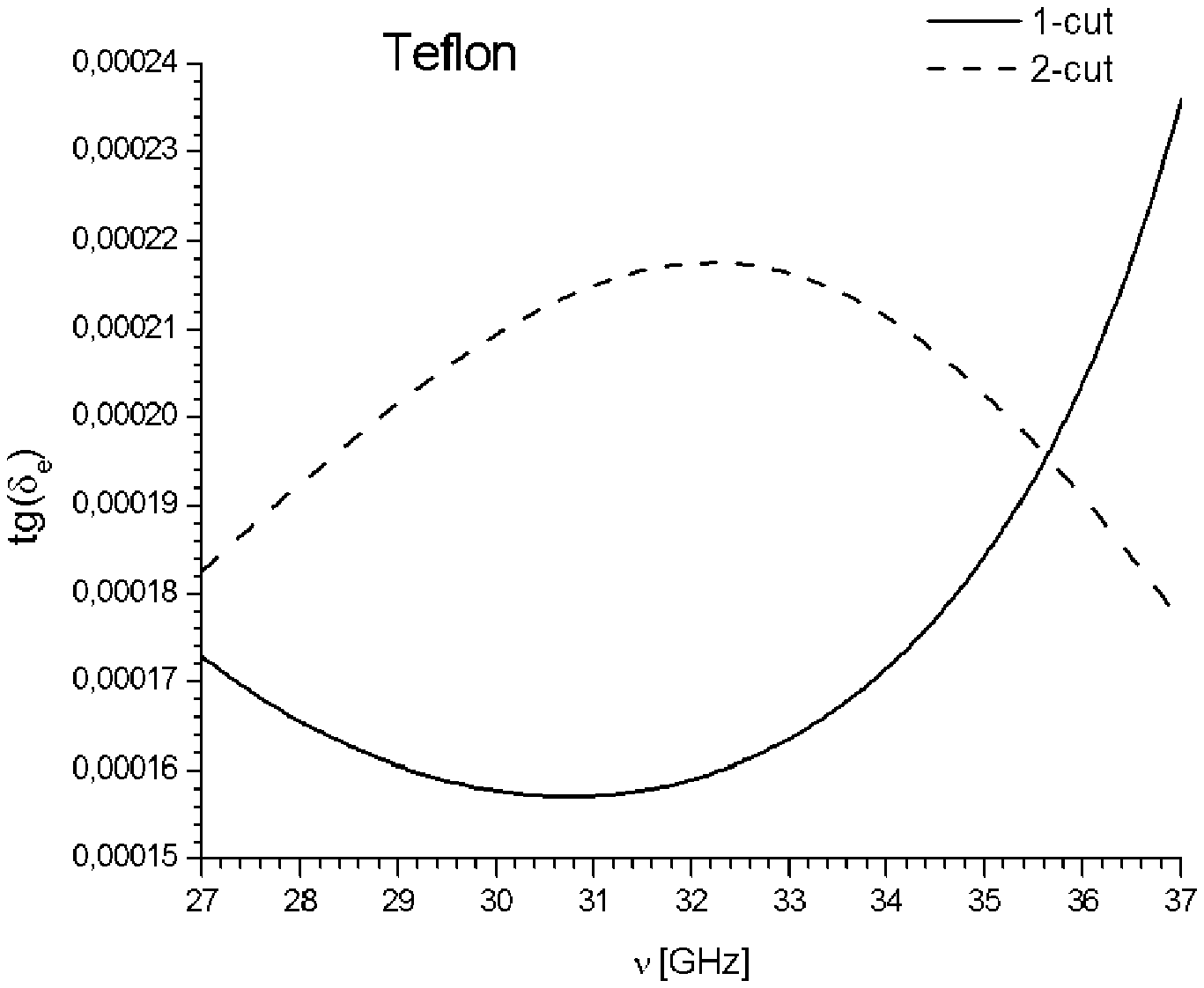}
        \\\\\\\\\\
    \includegraphics[width=2.7in,height=2.2in]{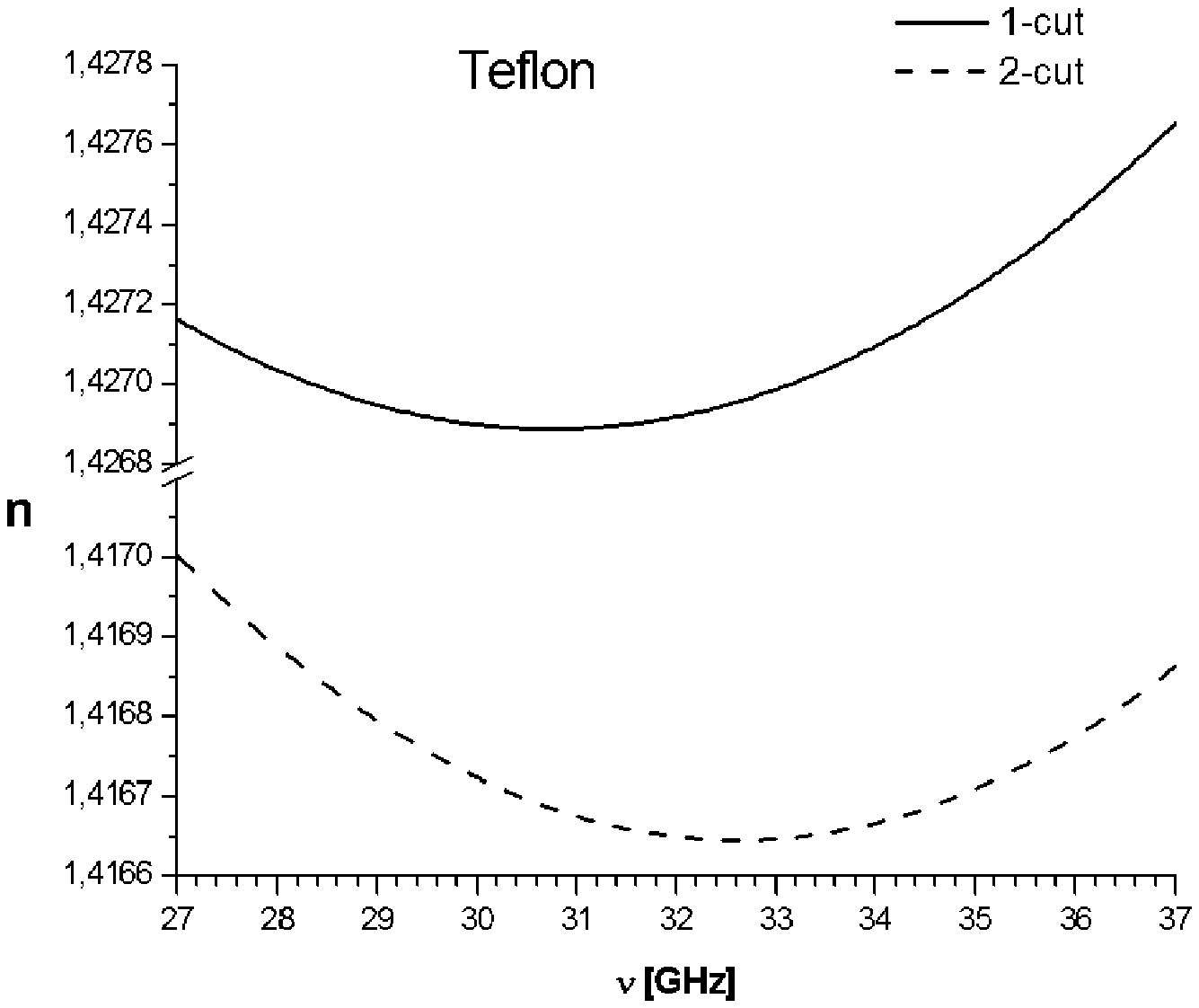} \hspace{15 mm}
    \includegraphics[width=2.7in,height=2.2in]{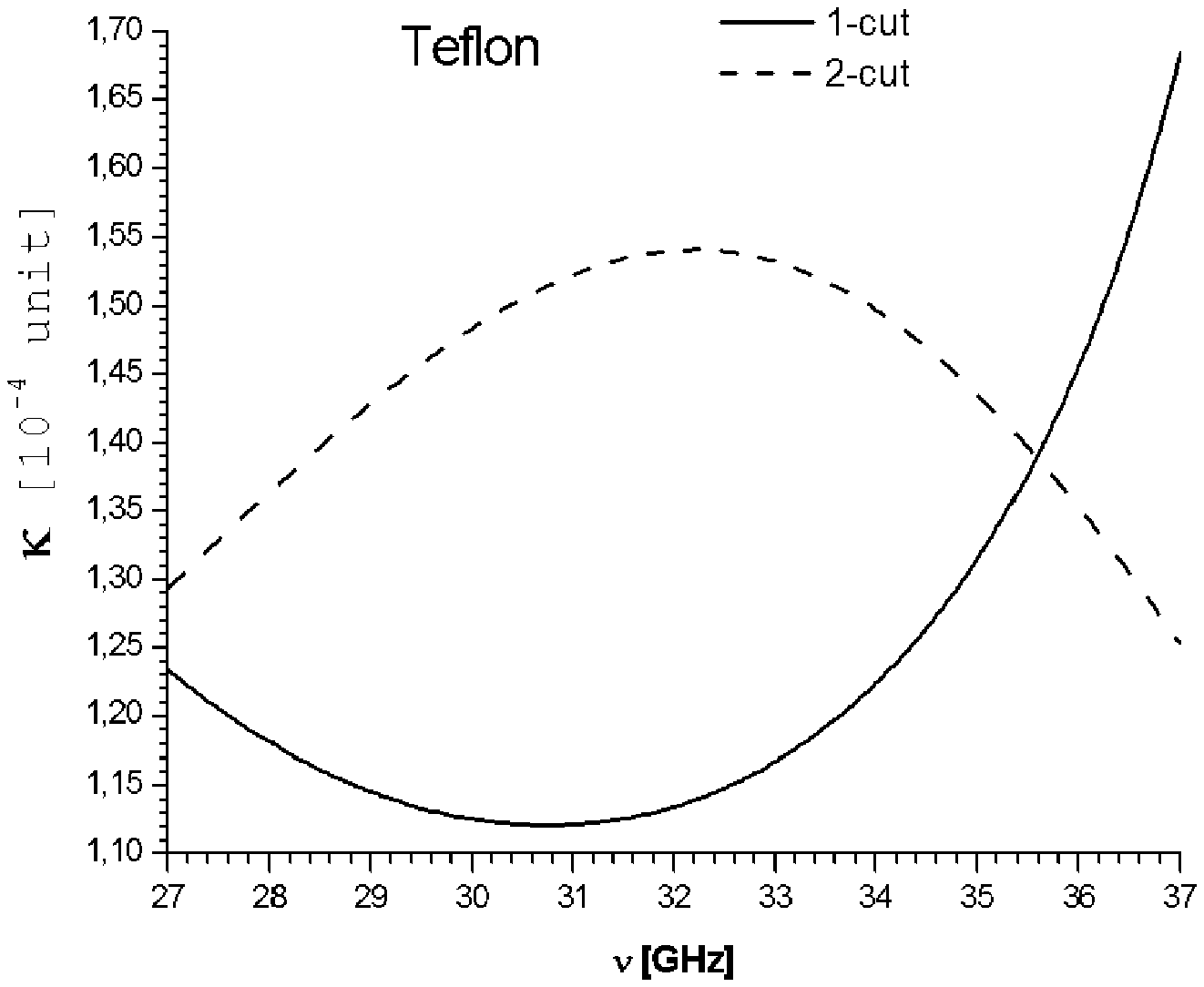}
    \caption{Complex dielectric constant and index of refraction of 1--cut
            and 2--cut Teflon. The measurements have been performed at
            T$_{\rm{ph}}$~=~300 K.  The precision for $\epsilon_r$
            and $n$ is $\sim$ 1$\%$, and for $\tan(\delta_e)$ and $\kappa$
            is $\sim$ 30$\%$.}
    \label{Teflon_anis_fig}
\end{figure}

In order to provide realistic estimates of the analysis hitherto
carried out, we performed measurements on several samples to
determine their complex dielectric constant $\bf{\epsilon}$ (${\bf
\epsilon} = \epsilon_{r}[1-i\tan(\delta_{e})]$, where
$\tan(\delta_{e}$) is the electric tangent loss), which entries in
the determination of the $R\!_{j}$, $T\!_{j}$ and
$\varepsilon\!_{j}$ quantities \cite{born,brehat,piro,azzam} by
means of the complex index of refraction $\textbf{n}$ ($\textbf{n}
= n-i \kappa$, where $\kappa$ is the extinction coefficient). In
fact, they are related as follow (for $\mu_{r} = 1$):
\begin{eqnarray}\label{dielectric_constant_refractive_index}
    n & = & \sqrt{\frac{\epsilon_{r}}{2}} \bigg [ \sqrt{1+\tan^{2}(\delta_{e})} + 1 \bigg
    ]^{1/2} \bigg |_{\;\tan(\delta_{e})<<1} \simeq \sqrt{\epsilon_{r}}\\
    \kappa & = & \sqrt{\frac{\epsilon_{r}}{2}} \bigg [ \sqrt{1+\tan^{2}(\delta_{e})} - 1 \bigg ]^{1/2}
    \bigg |_{\;\tan(\delta_{e})<<1} \simeq \frac{1}{2}\sqrt{\epsilon_{r}}\tan(\delta_{e})
\end{eqnarray}
Since electromagnetic properties of polymers can vary with the
composition, history and temperature of the specimen \cite{afsar},
measurements are mandatory. As a matter of fact, it is convenient
to extract the slab of interest from the same bulk of material
used to cut the samples under test.
\\
We have investigated High Density Polyethylene (HDPE), Teflon,
Polypropylene and Nylon, which are dielectrics commonly used from
microwave to far infrared frequencies. The tests have been
performed in the frequency band $27 < \nu < 37$~GHz, interesting
for microwave cosmology \cite{misha}. The measurements have been
performed by means of a vector network analyzer (Model HP8510). A
waveguide device has been realized in standard WR-28. It consists
of two shells cut along the E-plane (Fig.
(\ref{misure_dielettrici_foto})) which embeds the test sample.
The test is based on the comparison of the reflection and
transmission parameters $S_{11}$ and $S_{21}$ of the global device
\cite{baker} (waveguide plus sample), by making two measurements
with and without the dielectric test sample. The measured
insertion loss and return loss can be related to impedance and
propagation constants characterizing the equivalent transmission
line of the fundamental mode. In turn, these constants are related
to $\epsilon_{r}$ (one of the wanted quantity), the resistivity of
the material $\rho_{m}$ and the size of the sample, so allowing
the computation of the other relevant quantity,
$\tan(\delta_{e})$. Estimates of $\epsilon_{r}$ and
$\tan(\delta_{e})$ are carried out by performing a best fit on the
data and assuming a second order polynomial behavior for
$\epsilon_{r}$ and $\rho_{m}$. Such estimates will be specialized
by computing their in-band (30.4--33.6 GHz) average values, which
are those of interest to the BaR-SPOrt project. An example of our
results is reported in Fig. (\ref{simulation_data}), which shows
the measurements and the best fit model of the scattering
parameters in the case of a Teflon sample.
The main error source for the fitting procedure comes from the
knowledge of the sample sizes (reported in Table
(\ref{materials_size_table}); see also Fig.
(\ref{materials_size_fig}) for parameter definitions). Such errors
are taken into account when computing the in-band average values.
In Fig. (\ref{misure_dielettrici_1}) the complex constants are
shown while Table (\ref{optical_constant_risultati}) and
(\ref{in_band_var_risultati}) report the in--band average values
and variations.
For sake of completeness, we compare our data with those reported
in the literature (see Table (\ref{opt_constant_ref})). There is a
good consistency for the estimated dielectric constants. Large
differences, instead, are found for the electric tangent loss of
Teflon and Polypropylene. The poor precision is due to the
experimental technique adopted, which is not ideal for low loss
dielectrics. In this case, a more accurate evaluation of the
electric tangent loss could be obtained by applying the cavity
technique \cite{davide}. These considerations look supported by
the high precision data of both $\epsilon_r$ and $\tan(\delta_e)$
obtained for the Nylon sample. Although we provide just upper
limits for $\tan(\delta_e)$, no higher precision measurements seem
necessary since the contamination computed in Section
(\ref{application}) are already negligible.
Due to our interest in tiny polarization signal, we performed
extensive measurements on Teflon to investigate its optical
anisotropy. In fact, it seems to be more promising among the
materials selected. The manufacturing, in fact, can introduce in
these polymers anisotropic electromagnetic and structural
properties, since the molecular chains (structural units) will be
preferentially aligned in certain directions
\cite{polymer_1,polymer_2}, thus transforming the dielectric into
a sort of polarizer. Hence, we preferred to use a Teflon block
that has been casted and not extruded to minimize such
manufacturing effects. We assume such a block as homogeneous. By
considering the vibration of the electric field propagating in the
rectangular waveguide, it is possible to investigate the optical
anisotropy of dielectrics by cutting the sample as shown in Fig
(\ref{materials_size_fig}). Since typical feeds for microwave
cosmology are characterized by high directivity
\cite{polar,sport,nesti}, most of the signal will be collected
close to 0$^{\circ}$. Thus, two samples have been cut along the
z--axis, but rotated by 90$^\circ$ each other, to probe the
material for the incoming electric field vibrating in $x$ and $y$
directions.
\\
The results in the two directions are consistent each other within
the error (see Fig. (\ref{Teflon_anis_fig}) and Table
(\ref{Teflon_anis_table})), and possible anisotropy of the complex
index of refraction between $0^{\circ}$ and $90^{\circ}$ of
rotation is summarized by the following data:
\begin{eqnarray}\label{anisotropy_refractive_index}
    \Delta \langle n \rangle_\nu & = & \langle n \rangle_\nu^{\rm 1-cut} -
                                    \langle n \rangle_\nu^{\rm 2-cut} = 0.01 \pm 0.01 \\
    \Delta \langle \kappa \rangle_\nu & = & \langle \kappa \rangle_\nu^{\rm 1-cut} -
                                    \langle \kappa \rangle_\nu^{\rm 2-cut} = (4.2 \pm 5.8) \times 10^{-5}
\end{eqnarray}
which means that we can set at 1$\sigma$ of Confidence Level the
following upper limit: $\overline{\Delta \langle n \rangle_\nu} =
0.02$ and $\overline{\Delta \langle \kappa \rangle_\nu} =
10^{-4}$. These measurements allow us to consider our Teflon
sample optically isotropic at least for our purposes.
\section{Estimates of the systematic effects.}
\label{application}
\begin{figure} [h]
    \includegraphics[width=2.7in,height=2.2in]{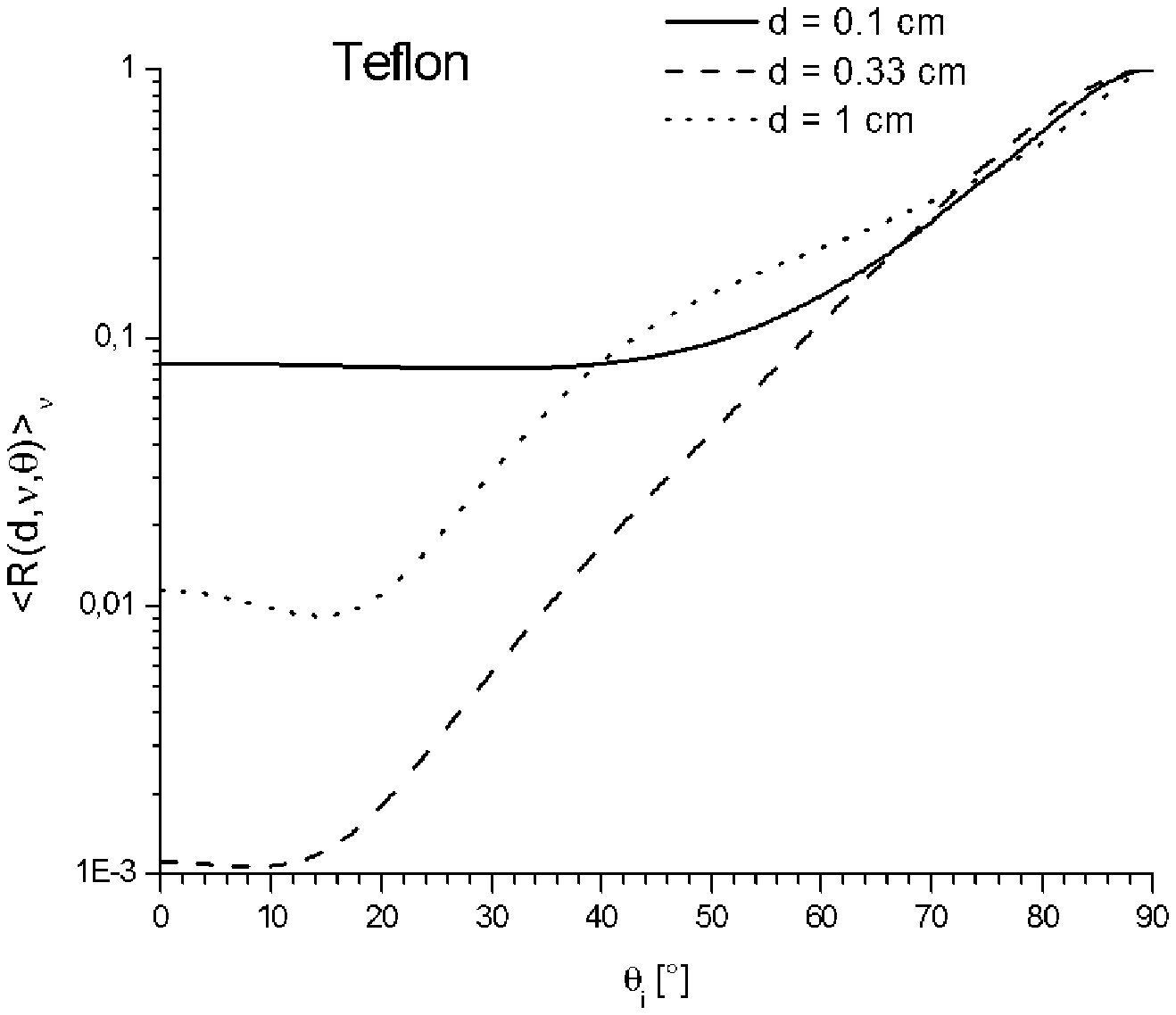} \hspace{15 mm}
    \includegraphics[width=2.7in,height=2.2in]{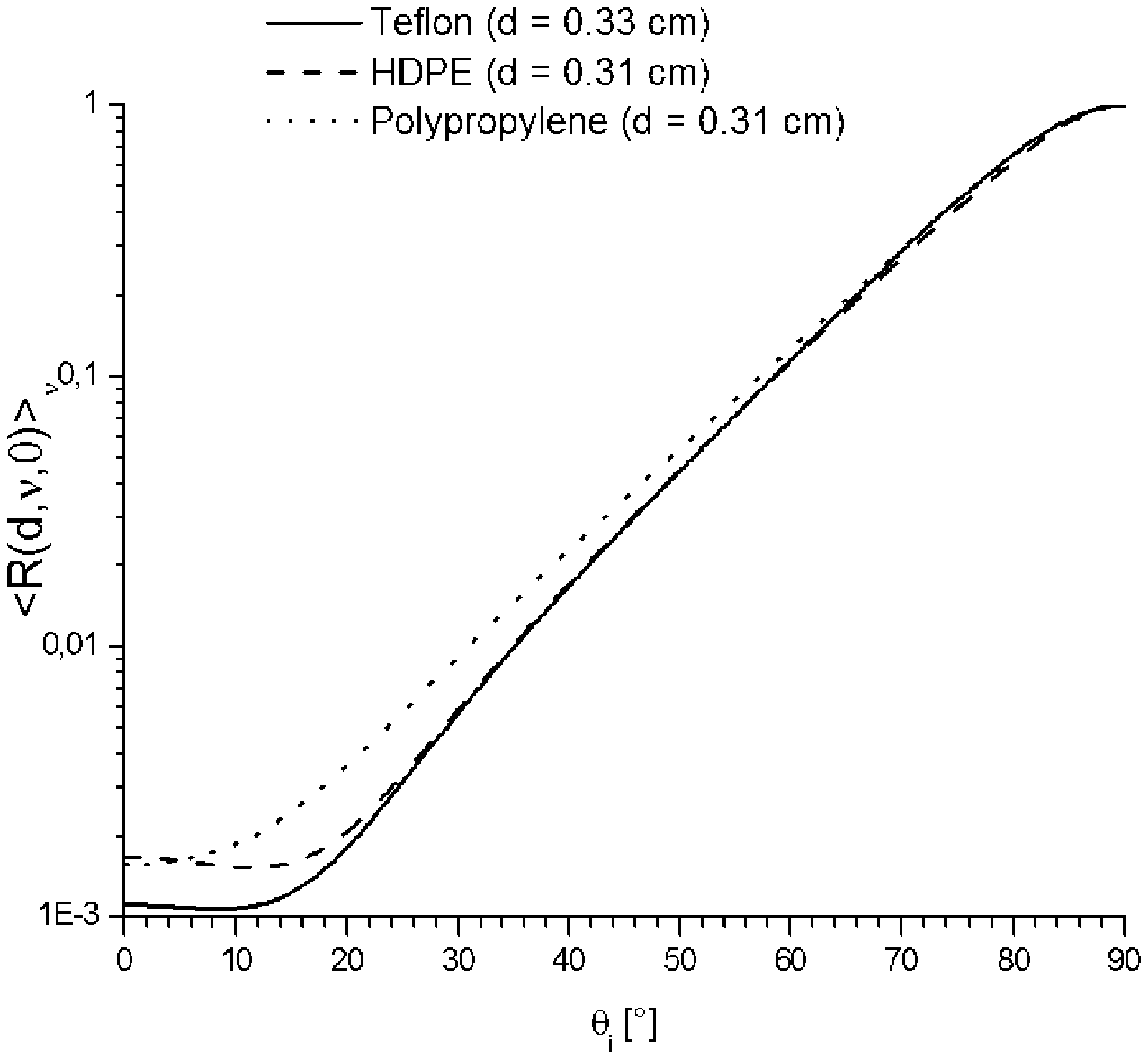}
    \\\\\\\\\\
    \includegraphics[width=2.8in,height=2.3in]{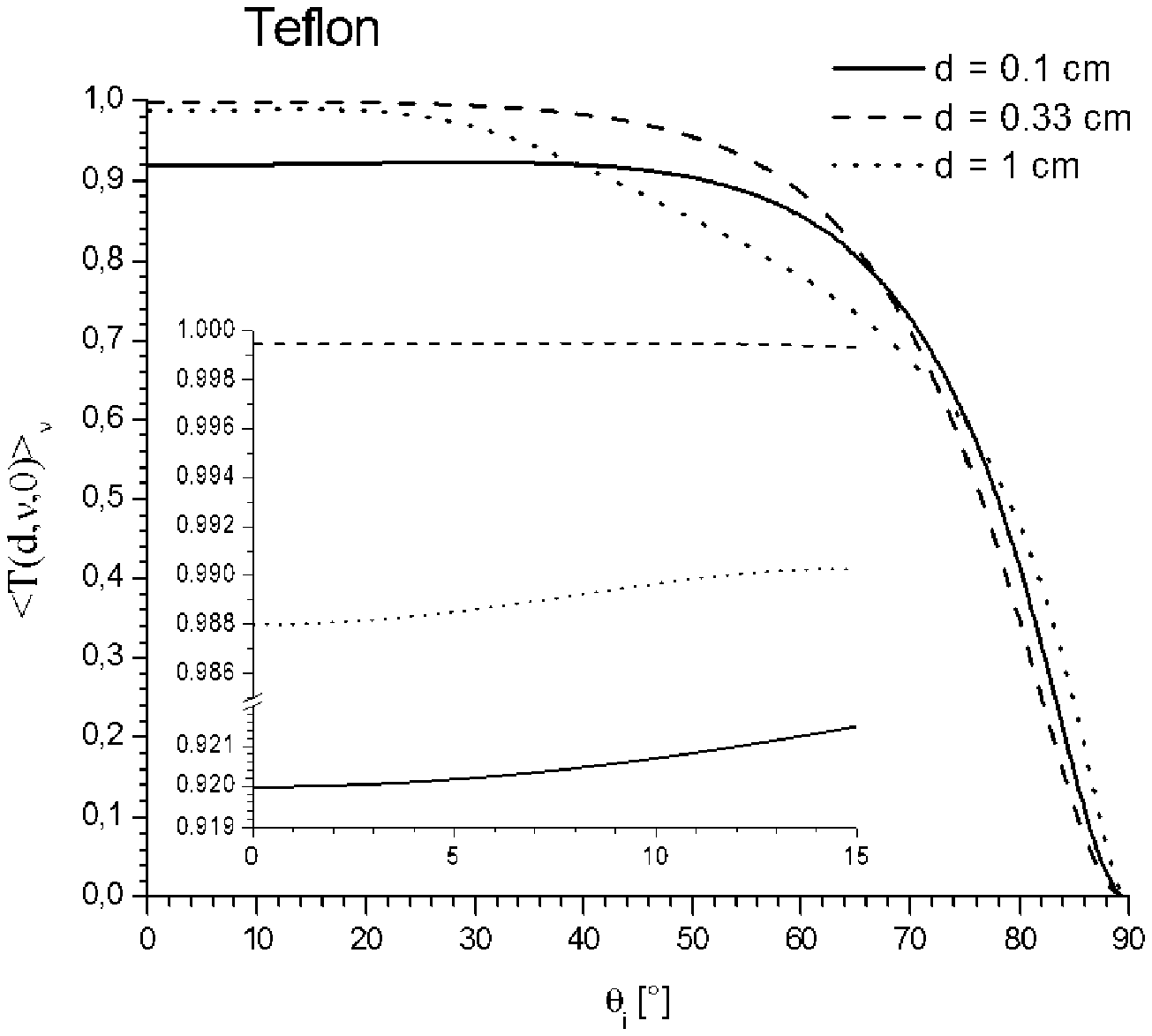} \hspace{14 mm}
    \includegraphics[width=2.8in,height=2.3in]{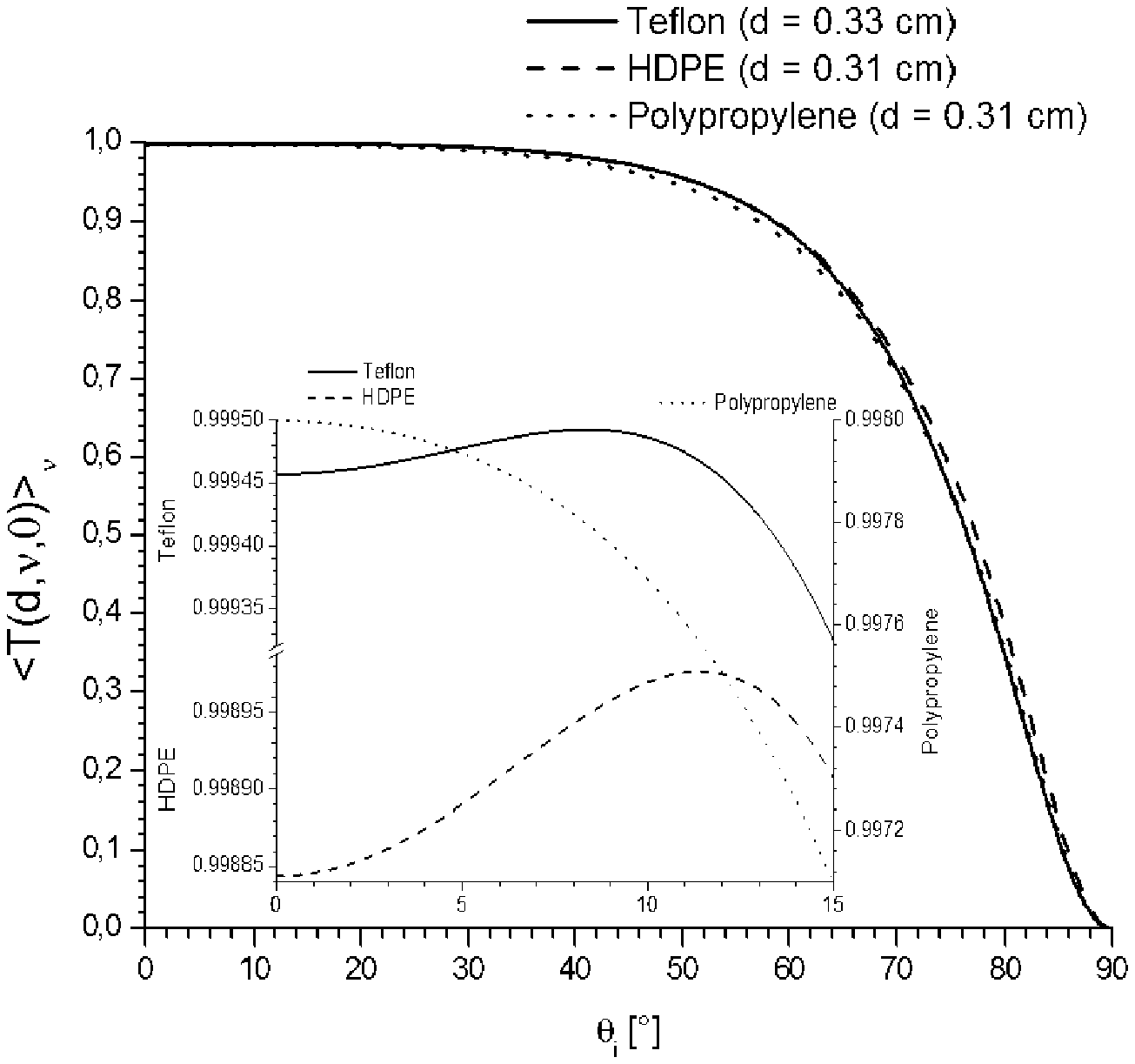}
    \\\\\\\\\\
    \includegraphics[width=2.7in,height=2.2in]{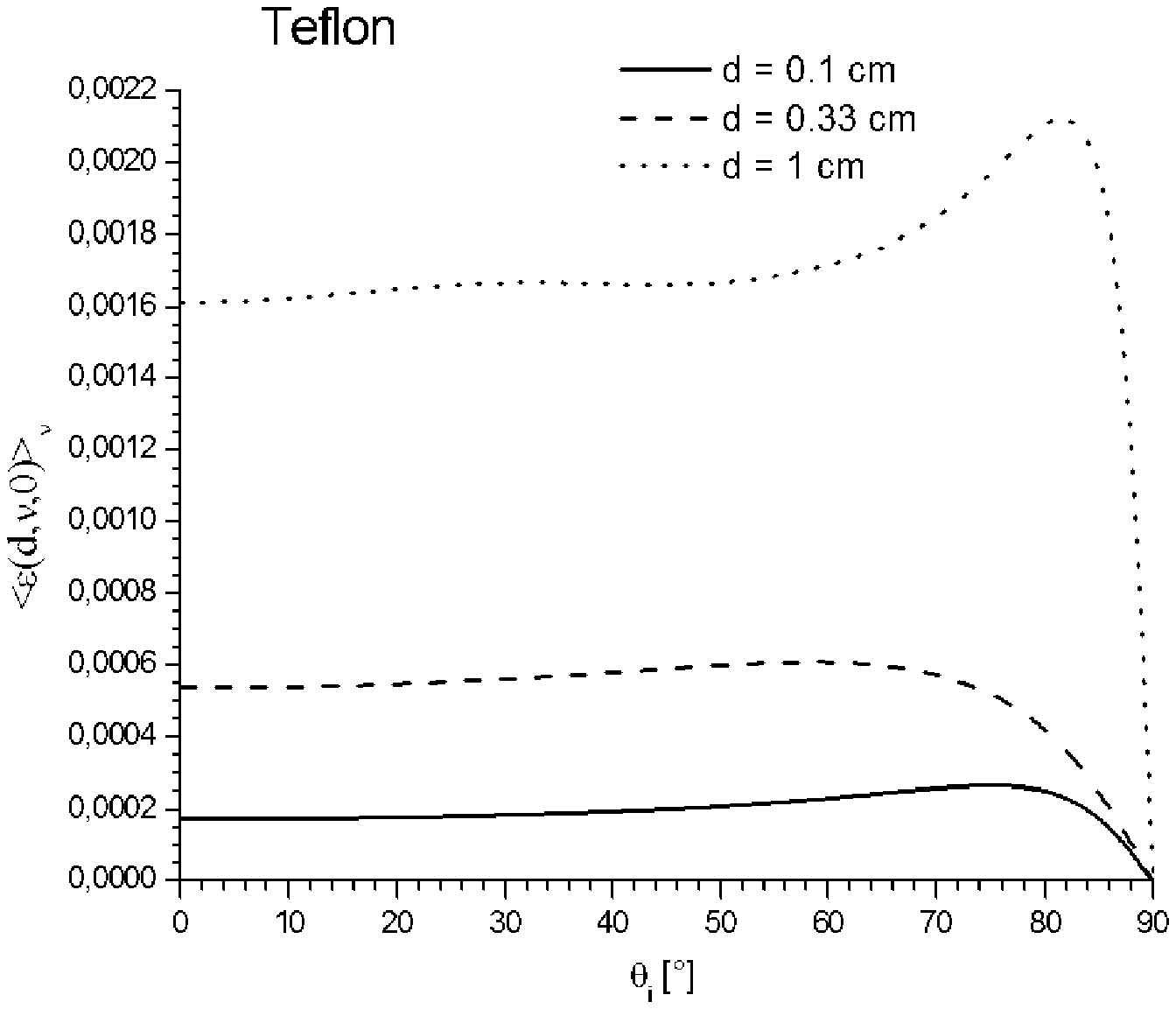} \hspace{15 mm}
    \includegraphics[width=2.7in,height=2.2in]{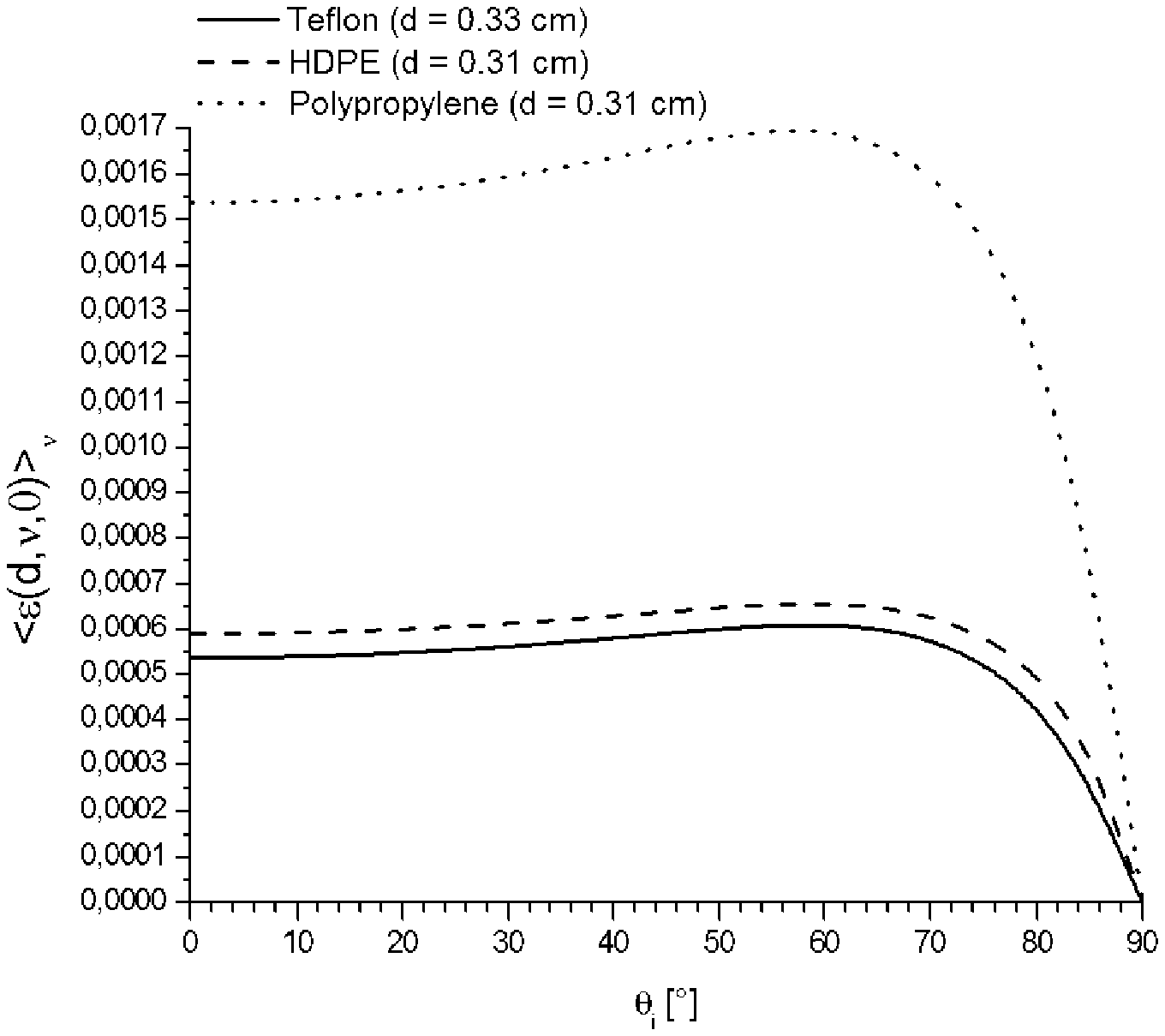}
    \caption{In band average reflectance, transmittance and
            emittance. Left: Teflon. Right: comparison among
            dielectrics. $d$ is the slab thickness.}
    \label{RTA_dielectrics}
\end{figure}

\begin{figure} [h]
    \centerline{\includegraphics[width=4in,height=3in]{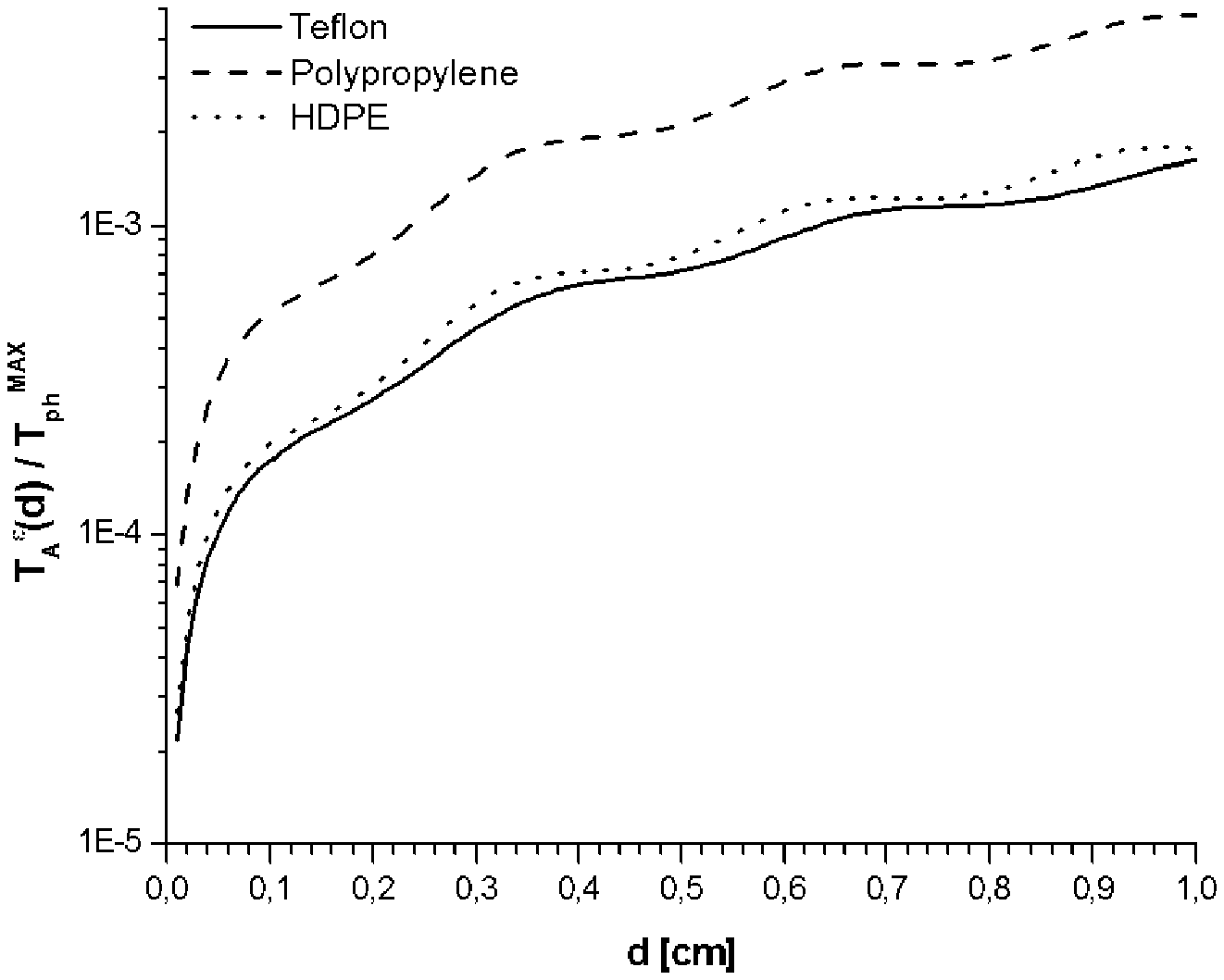}}
    \caption{BaR-SPOrt case. Top: the relative transmitted signal $\Lambda$.
            Bottom: upper limit of the relative emitted signal. $d$ is the slab thickness.}
    \label{TA_E_LAMBDA}
\end{figure}

\begin{figure} [h]
    \vspace{10 mm}
    \centerline{\includegraphics[width=3.5in,height=2.5in]{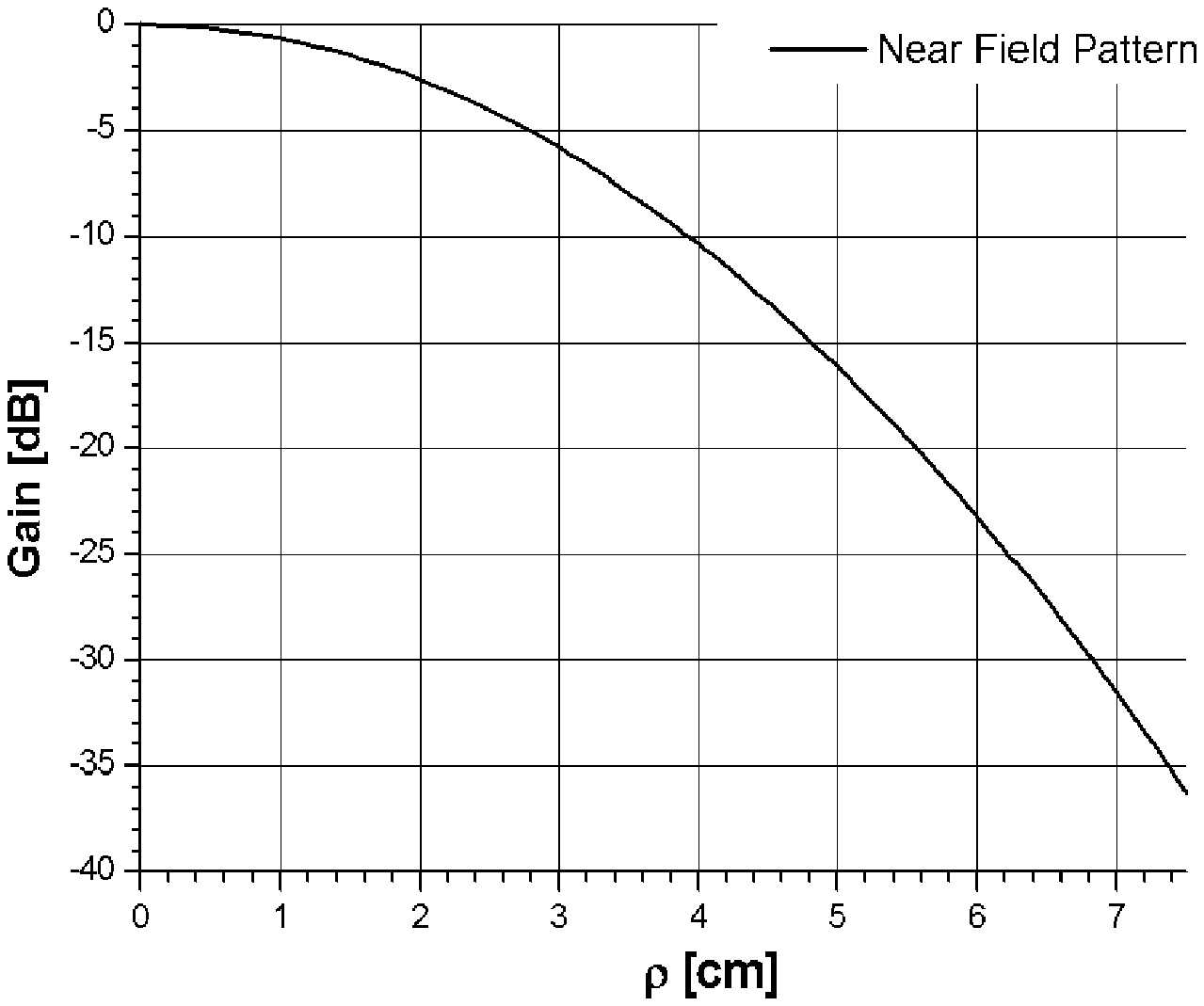}}
    \caption{Left: BaR-SPOrt far field patterns. Right: the near field versus the
            radial coordinate from the circular flat slab center ($\rho \leq {\rm R} = 7.5$
            cm, where R is the radius), by assuming a Gaussian beam approximation
            and using the following system parameters: distance between
            the feed aperture and the slab ($\delta$~=~6.5 mm); confocal
            distance ($z_{c}$ = 204.56 mm); waist ($w_{0}$ = 24.707~mm)
            and the distance between the waist and the slab ($z = 224.716$ mm)
            (see Ref. \citeonline{nesti, nesti_private}).}
    \label{Polar_pattern}
\end{figure}

\begin{figure} [h]
    \includegraphics[width=2.7in,height=2.2in]{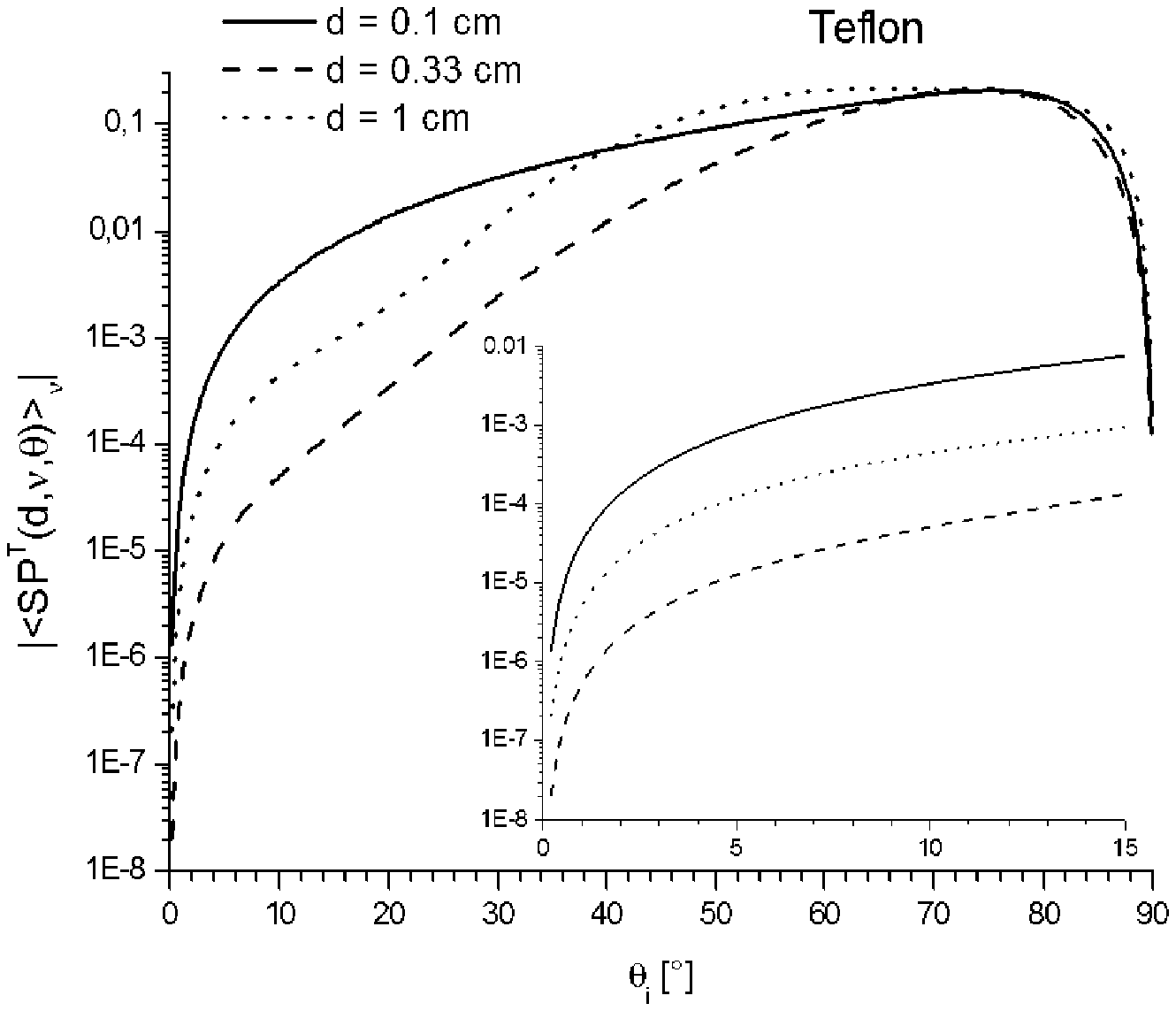} \hspace{15 mm}
    \includegraphics[width=2.7in,height=2.2in]{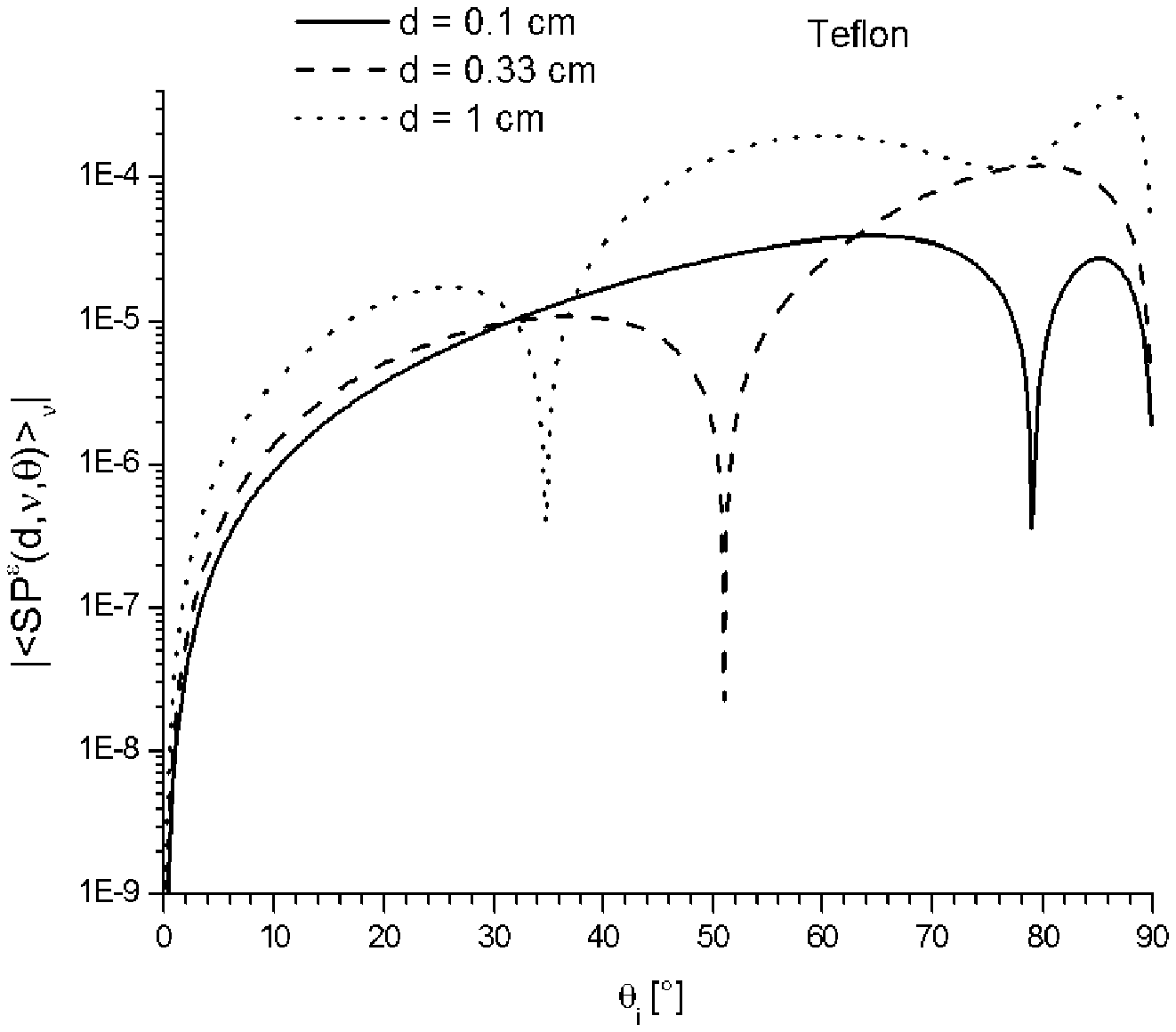}
    \\\\\\\\\\
    \includegraphics[width=2.7in,height=2.2in]{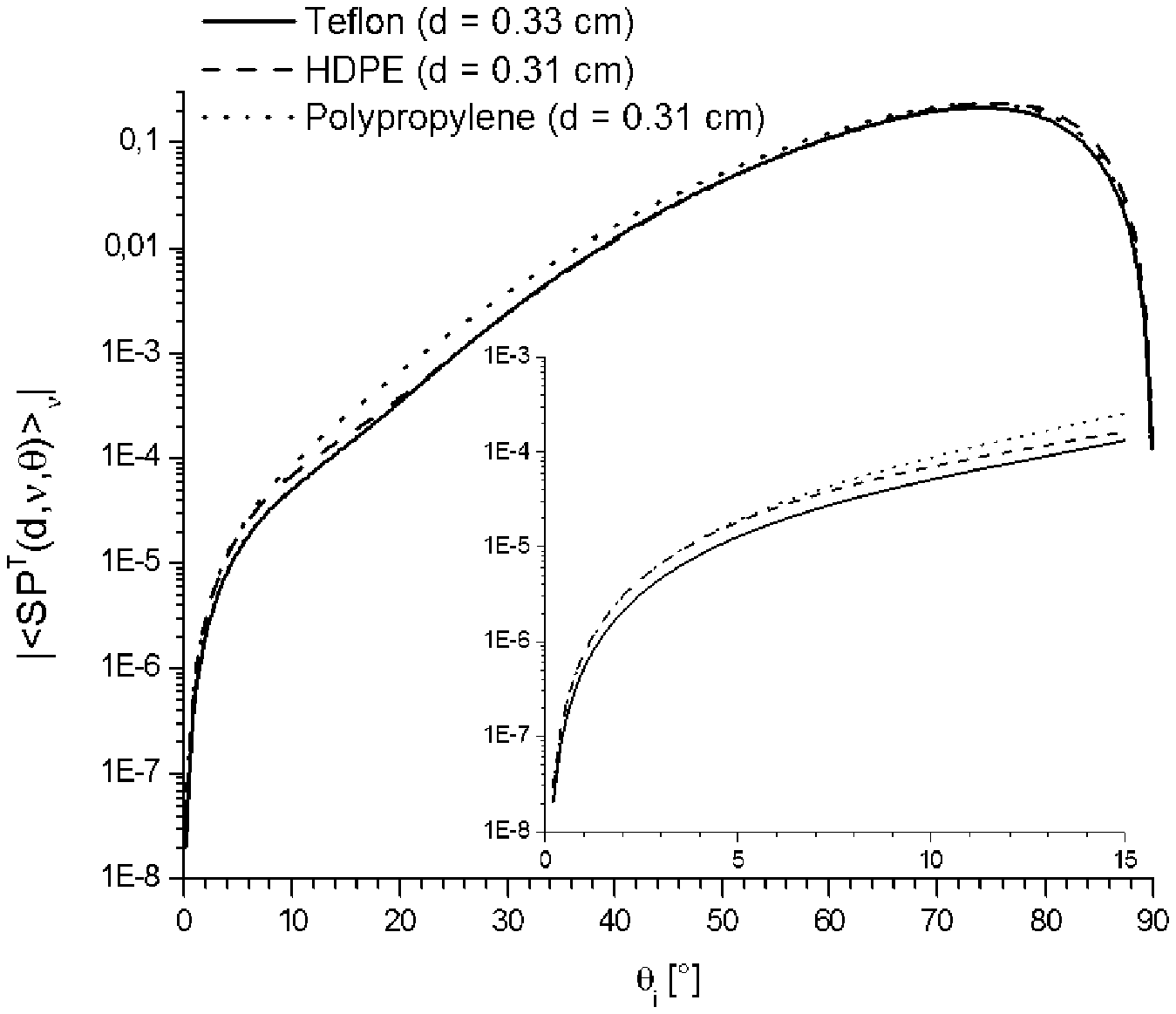} \hspace{15 mm}
    \includegraphics[width=2.7in,height=2.2in]{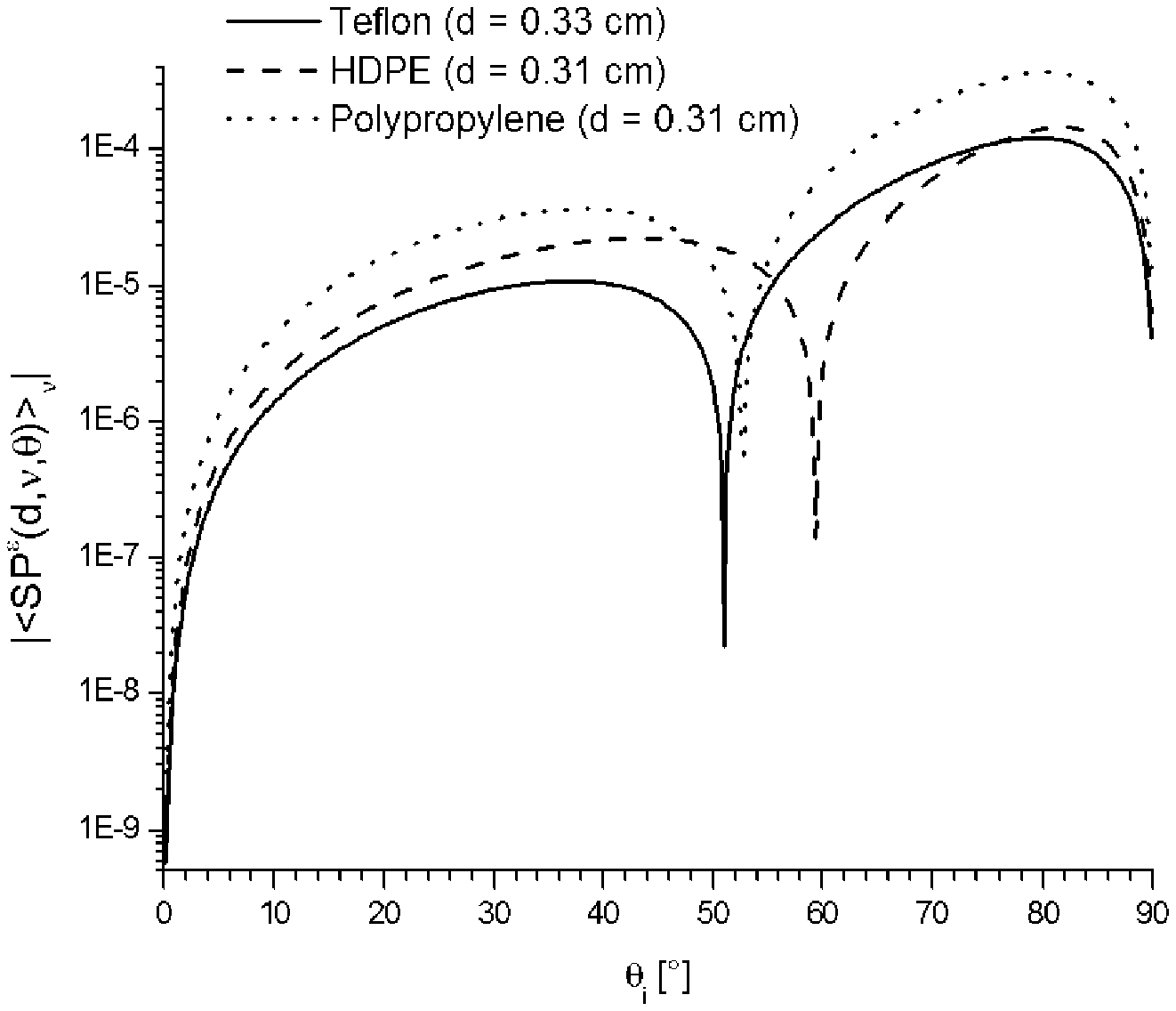}
    \caption{Absolute value of the spurious polarization coefficients related to Teflon, HDPE and Polypropylene
            in the 30.4--33.6 GHz band. Left: transmission regime. Right: emission regime.
            $d$ is the slab thickness.}
    \label{spurious_coefficient_isotropic_fig}
\end{figure}

\begin{figure} [h]
    \vspace{10 mm}
        \centerline{\includegraphics[width=3.3in,height=2.3in]{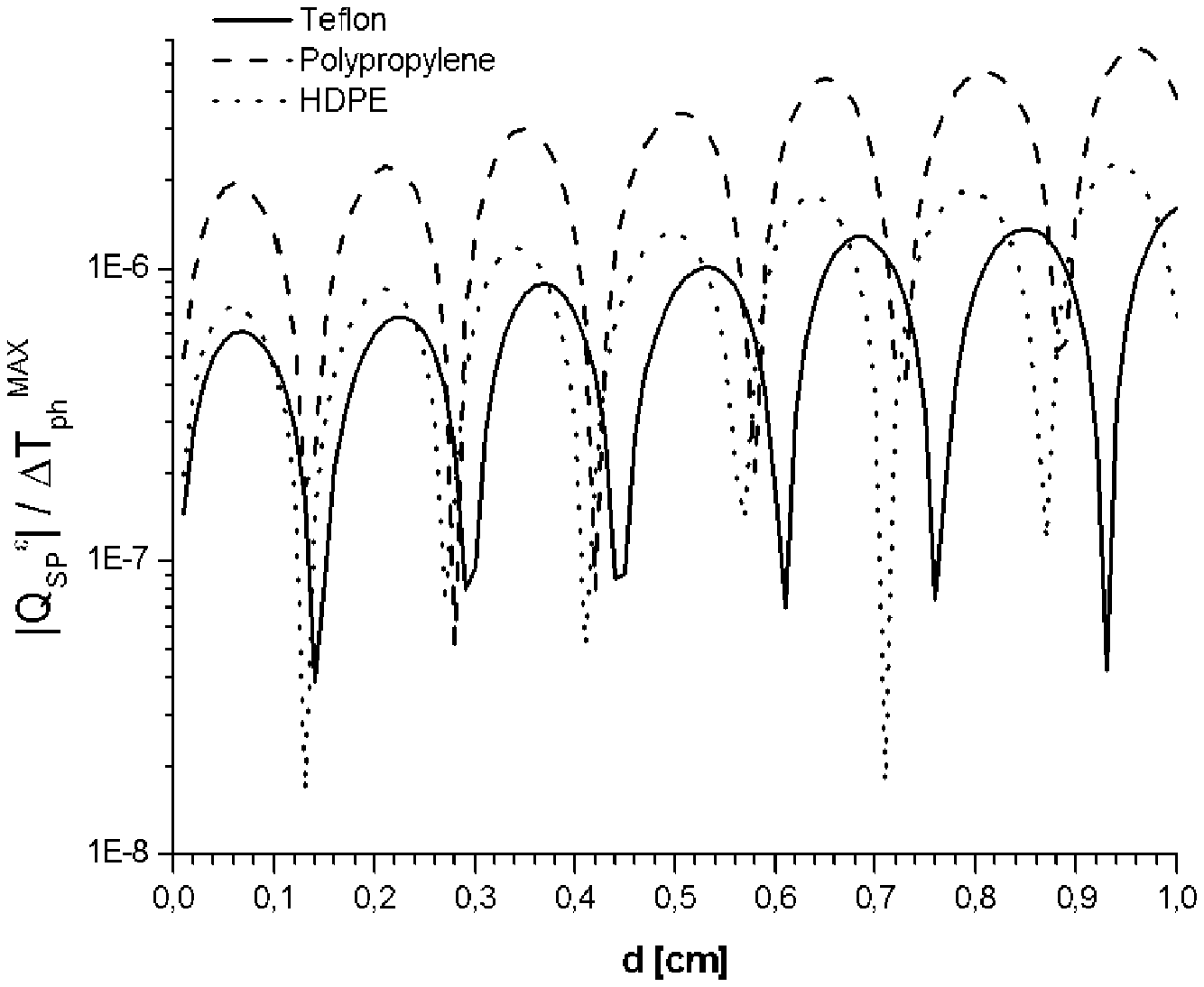}}
    \caption{BaR-SPOrt case. Top: upper limit of the spurious polarized transmission.
            Bottom: upper limit of the spurious polarized emission. (Similarly for $U$).
            $d$ is the slab thickness.}
    \label{SPT_SPE_ant_temp}
\end{figure}

\begin{figure} [h]
    \vspace{15 mm}
    \centerline{\includegraphics[width=3.3in,height=2.4in]{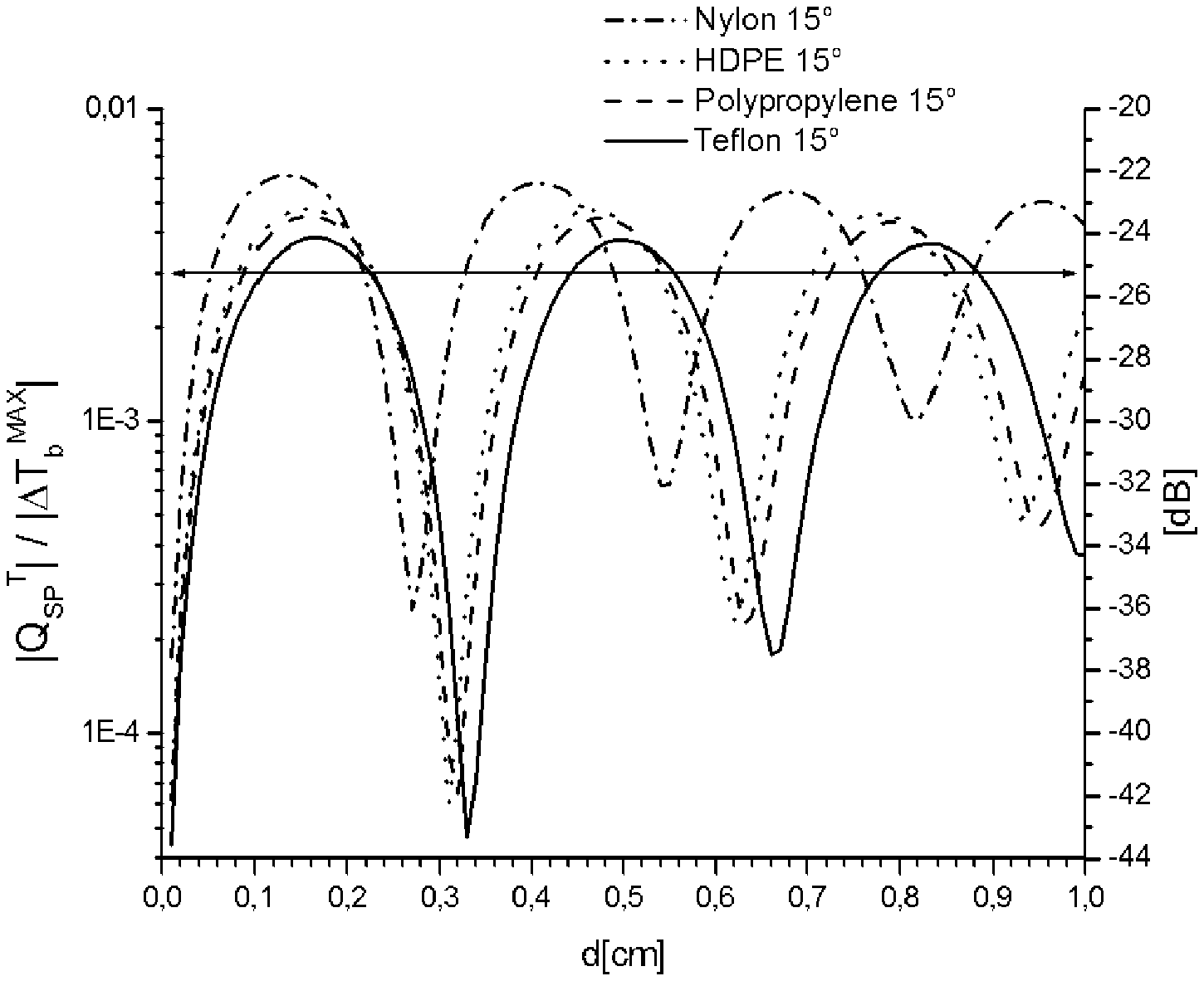}}
    \caption{Plots of the spurious polarization transmitted by flat dielectrics in
            far field regime, in the 30.4--33.6 GHz band, by
            increasing the beam pattern. Nylon is shown only for
            comparison. $d$ is the slab thickness. See text for details.}
    \label{SPT_beam_fig}
\end{figure}

\begin{figure} [h]
    \vspace{10 mm}
    \centerline{\includegraphics[width=3.4in,height=2.4in]{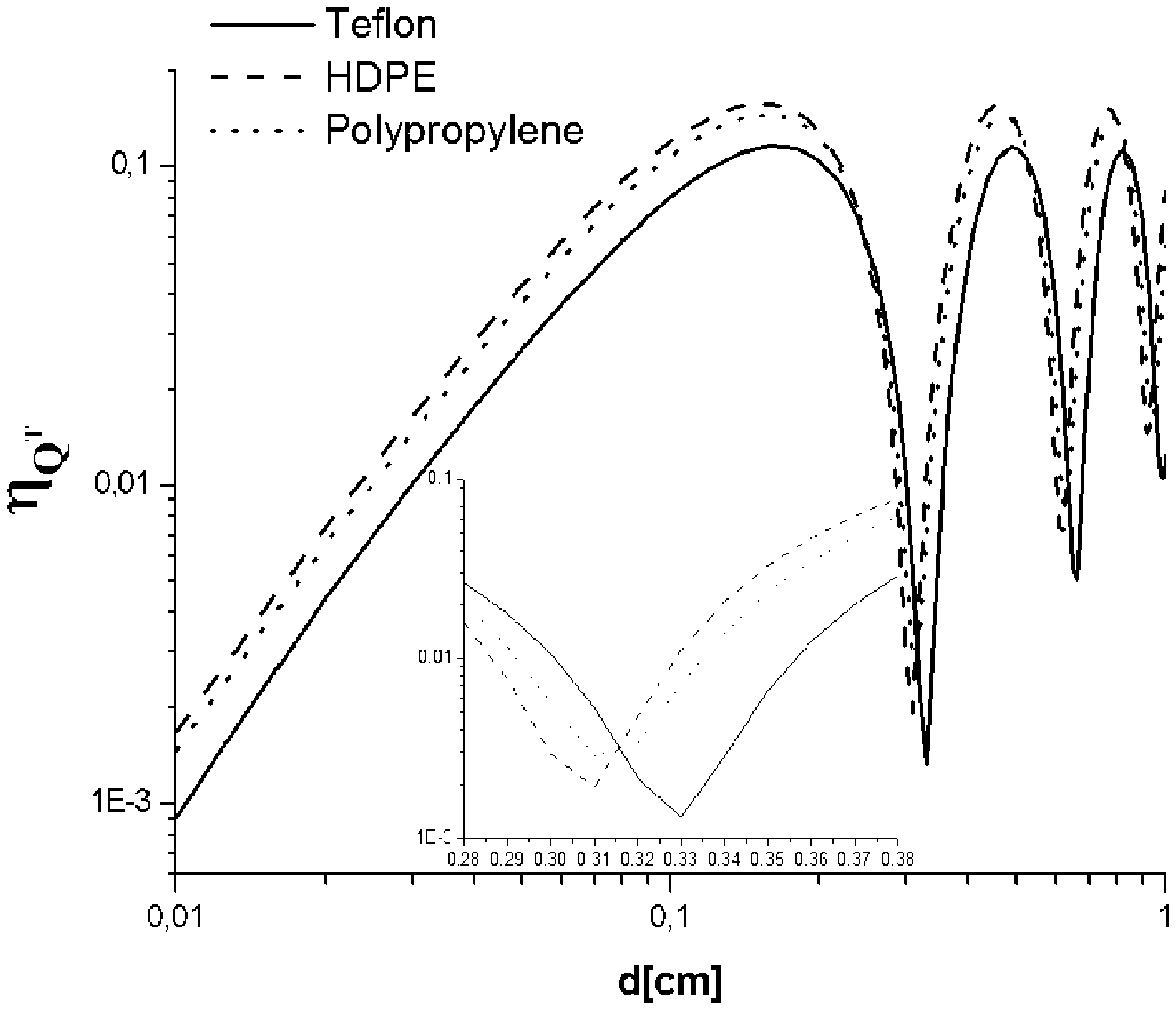}}
    \caption{Plots of the $\tilde{D}_{Q}^{T}$ term (top) and of the depolarization
            in transmission regime (bottom) introduced by Teflon, HDPE and
            Polypropylene in the 30.4--33.6 GHz band. $d$ is the slab thickness.}
    \label{depolarization_isotropic_fig}
\end{figure}

\begin{figure} [h]
    \includegraphics[width=2.7in,height=2.2in]{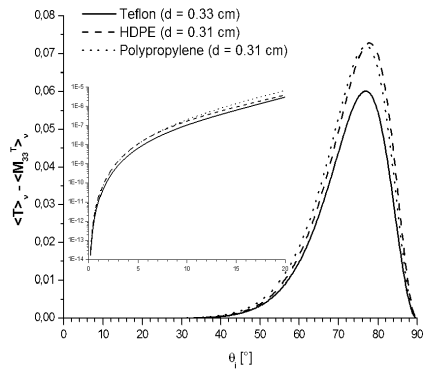} \hspace{12 mm}
    \includegraphics[width=1.45in,height=1.2in]{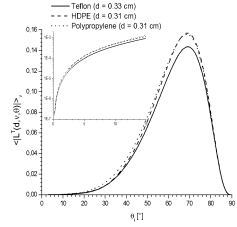}
    \\\\\\\\\\
    \includegraphics[width=2.7in,height=2.2in]{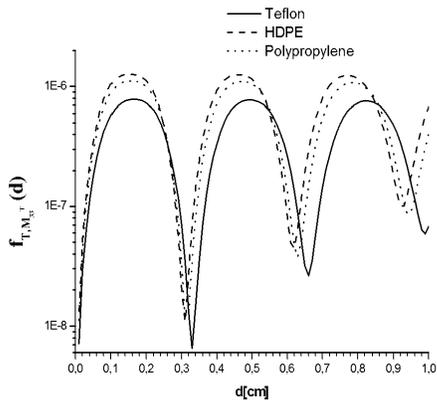} \hspace{15 mm}
    \includegraphics[width=1.25in,height=1in]{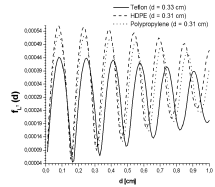}
    \caption{Plots of the leakage effects for Teflon, HDPE and Polypropylene
            in the 30.4--33.6 GHz band. Bottom plots are for the BaR--SPOrt case.
            $d$ is the slab thickness.}
    \label{leakage_isotropic_fig}
\end{figure}

The analysis performed in Section (\ref{is_diel_sp_pol}) includes
the far and near field regime, even though for our purpose most
estimates will be provided for a flat slab placed in near field
position. However one case of far field will be also considered.
We take into account an instrument featured by a $10 \%$
bandwidth, typical of recent microwave polarimeters, so that all
the relevant quantities of Sections (\ref{is_diel_sp_pol}) and
(\ref{is_diel}) are evaluated as in-band average. Furthermore,
here we replace the angle $\theta^{\; \prime}$ with $\theta$ (i.e.
the incoming Far Field direction) making easier the computation of
all the interesting quantities. This approximation does not
prevent the aim of this work. In fact, if the slab is close to the
feed aperture then $\theta^{\; \prime} < \theta$, and, in turn,
this substitution provides conservative estimates. The materials
considered here are Teflon, HDPE and Polypropylene, disregarding
the Nylon due to its high values of $\epsilon_{r}$ and
$\tan(\delta_{e})$, since we are interested in the minimization
analysis of the systematic effects. For the Teflon, we take into
account the 1--cut, since its complex index of refraction is more
precise than the other~cut.
\\
In Fig. (\ref{RTA_dielectrics}) the in-band average reflectance,
transmittance and emittance of the dielectrics versus $\theta_{i}$
are shown for 3 thickness values. For each panel the central value
is the thickness which maximizes the transmission at $\theta_i =
0$ (see the Teflon plots): it corresponds to $d = d_{0} \equiv
\lambda/2n$ where $\lambda \sim 0.94$ cm. As expected, these plots
show that either increasing $\kappa$ or the thickness generates an
increase of the emittance. Similarly, the reflectance increases by
increasing $n$. Close to the axis, transmittance variations are
very small (for $\theta_{i} < 15^{\circ}$ is lower than
$10^{-3}$), then $T(d,\nu,\theta^{\; \prime}) \simeq
T(d,\nu,\theta)$, so allowing the substitution $\theta^{\;
\prime}$ with $\theta$. In these frequencies such dielectrics show
very low losses ($\sim$ $10^{-3}\div10^{-4}$, see the emittance
plot), then the thickness $d_{0}$ provides a good estimate of the
thickness that maximizes the transmittance.
\\
A first estimate of the thermal noise (${\rm T}_{\rm noise}$)
injected by the slab can be given by Eq.
(\ref{emiss_out_sovrastima}) using the parameter $\langle
\varepsilon(d,\nu,\theta) \rangle_{\nu}|_{{\rm MAX}(\theta)}$ (see
bottom-right plot of Fig. (\ref{RTA_dielectrics}) and Table
(\ref{noise_temp_dielectrics}) for the estimates).
\\
A better estimate of the signal transmission and of the thermal
noise collected by the feed (Fig. (\ref{TA_E_LAMBDA})) can be
given using Eq. (\ref{trans_out_e}) and (\ref{emiss_out}) where,
for the emission, the near field pattern has to be used. The
adopted patterns are shown in Fig. (\ref{Polar_pattern}), by using
the BaR-SPOrt far field one as realistic example \cite{nesti}, and
by assuming a Gaussian beam approximation for the near field
\cite{goldsmith}. In the last case, it is necessary to set a
geometrical configuration of the slab in front of the feed
aperture to produce the near field pattern. Then, for our purpose,
a circular flat slab will be adopted. Hereafter, we will refer to
the estimates of quantities related to antenna integrals as
``BaR-SPOrt case'' but, as we will show in this section, the core
idea can be applicable in the same way to other cases.
The plot in Fig. (\ref{TA_E_LAMBDA}) shows, as expected, that the
noise due to the slab increases with the thickness. The plot of
the relative transmitted signal $\Lambda$ shows the typical
interference trend. As highlighted in the magnified frame, the
value of the first maximum correspond to a thickness of $0.33$ cm
for the Teflon and $0.31$ cm for HDPE and Polypropylene, that is
the ideal value $d_{0}$ in case of the low absorbing material
approximation. It is due to the narrow beam of the feed adopted
which favors angles very close to $0^{\circ}$. Thus, in case of
narrow far field pattern, such a thickness provides the best size
for maximizing the transmittance of the slab. By choosing as
optimal thickness $d_{0}$, the residuals with respect to 1 (i.e.
$\Lambda(d) - 1$) give an estimate of the signal loss ($\sim
10^{-3}$). Besides the slab thickness optimization, this analysis
allows us to choose even the best material, which in the band
adopted is Teflon.
\\
Table (\ref{noise_temp_dielectrics}) shows the parameters $\langle
\varepsilon(d,\nu,\theta) \rangle_{\nu}|_{{\rm MAX}(\theta)}$ and
the quantity T$_{A}^{{\rm \varepsilon}}(d)/$T$_{{\rm ph}}^{{\rm
MAX}}$ obtained by increasing the antenna integral in Eq.
(\ref{emiss_out}) that are used to evaluate the noise temperature
injected by the slab (T$_{\rm noise}$). Between the two estimates,
the beam integration gives a correction of $\sim$ 10\% reducing
the value provided by the Eq. (\ref{emiss_out_sovrastima}): the
estimates are the same within 10\% variation. Thus, the Eq.
(\ref{emiss_out_sovrastima}) is an easy way to estimate the
thermal noise. As shown, the Teflon is the material which injects
the lowest noise (162 mK at T$_{{\rm ph}}^{{\rm MAX}} = 300$ K),
even though HDPE is just slightly worst and can be considered as
well.
\\
The spurious polarization coefficients are shown in Fig.
(\ref{spurious_coefficient_isotropic_fig}). Oscillations in the
spurious response are generated by increasing the optical path
inside the slab as is typical for interference phenomena. Once
again, the approximation for the angle can be applied since, close
to the feed axis, the relation $| \langle
S\!P^{T}(d,\nu,\theta^{\; \prime}) \rangle_{\nu}| < | \langle
S\!P^{T}(d,\nu,\theta) \rangle_{\nu}|$ holds. Then a conservative
estimate of Eq. (\ref{stokes_30})--(\ref{stokes_36_bis}) can be
represented by:
\begin{equation}\label{over_est}
    Q^{x}_{S\!P}|_{{\rm MAX}} = U^{x}_{S\!P}|_{{\rm MAX}} = |\langle S\!P^{x} (d,\nu,\theta)
            \rangle_{\nu} |_{{\rm MAX} (\theta)} \times |\Delta \! {\rm T}(\nu,\theta,\beta)|_{{\rm MAX} (\nu,\theta,\beta)}
\end{equation}
where $x$ is either $T$ or $\varepsilon$, and $\Delta \! {\rm T}$
is the variation of either the brightness or the physical
temperature.
\\
The upper limit of the spurious polarized transmission and
emission are plotted in Fig. (\ref{SPT_SPE_ant_temp}) for the
BaR-SPOrt case by conservatively approximating the Eq.
(\ref{stokes_30}) as follows:
{\setlength\arraycolsep{0pt}
\begin{eqnarray}
    Q_{\rm S\!P}^{T}(d) & = & \frac{1}{\Omega_{A}^{\rm FF}\Delta\nu} \int_{\Delta \nu} \textrm{d}\nu
                     \int_{0}^{\pi/2}\textrm{d}\theta S\!P^{T}(d,\nu,\theta^{\;\prime})P_{n}^{\rm FF}(\theta)
                    \sin(\theta) \nonumber \times \\
                    & \times & \int_{0}^{2 \pi}\textrm{d}\beta \Delta \! {\rm T\!}_{\rm b}(\nu,\theta,\beta)\cos(2\beta) \\\label{major_stokes_30_bis}
                    & < & |\Delta \! {\rm T\!}_{\rm b}(\nu,\theta,\beta)|_{{\rm MAX}} \;\; \frac{2 \pi}{\Omega_{A}^{\rm FF}}
                     \int_{0}^{\pi/2} |\langle S\!P^{T}(d,\nu,\theta) \rangle_{\nu}| P_{n}^{\rm FF}(\theta)
                    \sin(\theta) \; \textrm{d}\theta
\end{eqnarray}}
while the Eq. (\ref{stokes_36}), computed in cylindrical
coordinates, has been approximated as:
\begin{equation}\label{major_spe}
    Q_{\rm S\!P}^{\varepsilon}(d,z)
                    < \Delta \! {\rm T\!}_{\rm ph}(\nu,\rho,\beta)|_{{\rm MAX}} \;\; \frac{2 \pi}{\Omega_{A}^{\rm NF}}
                    \int_{0}^{R} \; |\langle S\!P^{\varepsilon}(d,\nu,\rho) \rangle_{\nu}| \; P_{n}^{\rm NF}(\rho,z)
                    \;\; \rho \; \textrm{d}\rho
\end{equation}
where $R$ is the radius of the circular slab and $z$ its position
with respect to the feed horn waist.
Such plots show that the low relative spurious polarization of the
transmitted component ($10^{-3} \div 10^{-5}$) is minimized by the
same thickness $d_{0}$ which maximizes the transmittance. For the
emitted component, the integrated spurious term is very low
($10^{-6} \div 10^{-8}$).
\\
Once again, it could be useful to compare the two methods to
evaluate the spurious polarization. The first one is defined by
the Eq. (\ref{over_est}), while the second one by Eq.
(\ref{major_stokes_30_bis}) and (\ref{major_spe}), which even
accounts for the antenna pattern. Results, shown in Tables
(\ref{sp_finestra_isotropa}) and (\ref{sp_finestra_isotropa_2}),
are computed by taking into account as example the known
anisotropic signal of the CMB ($\Delta \! \rm{T}_{b}^{{\rm MAX}} =
100 \mu$K) and a guess value of $\Delta \! \rm{T}_{\rm{ph}}^{{\rm
MAX}} = 1$ K related to the thermal gradient of the slab.
\\
Taking into account the antenna pattern, in the BaR-SPOrt case
such values are $\sim$ 200 times lower than the rough estimate
provided by the Eq. (\ref{over_est}), showing how critical it is
to consider the antenna pattern in this case. This is due to the
high directivity of the feed horn adopted and to the rapid
decrease of the $|S\!P|$ functions close to $0^{\circ}$. Teflon is
the material which introduces the lowest spurious polarizations
($\sim 0.6 \mu$K).
\\
For sake of completeness, we insert an analysis of the spurious
polarization generated in transmission regime as the beam of the
collecting system increases (Fig. (\ref{SPT_beam_fig})) when the
slab is placed in far field (i.e. $\theta^{\; \prime} \equiv
\theta$). Here a Gaussian pattern has been considered, thus
performing a general but optimistic analysis, since in case of
co-polar pattern of real feed the expected spurious signal is
greater than the level generated by a Gaussian one. In fact, the
Gaussian pattern shows a rapid decrease out of the beam, thus
producing lowest response as shown for comparison between the
minima of Teflon at 7$^{\circ}$ in the left-plot of Fig.
(\ref{SPT_beam_fig}) and in Fig. (\ref{SPT_SPE_ant_temp}) (the
difference is a factor 3, $\sim$ 5 dB).
The plots in Fig. (\ref{SPT_beam_fig}) show that the broader the
beam the higher the spurious level generated by the slab. As
expected, this is due to the rapid increase of the spurious
coefficient $|S\!P^{T}|$ out of the null incidence (see Fig.
(\ref{spurious_coefficient_isotropic_fig})). Moreover, the
thickness setting positions of minima varies with the beam due to
the low directivity of the collecting system as the beam
increases. In particular, such positions are different from those
expected for narrow beam around null incidence (e.g. see the
vertical lines across the left-plot of Fig. (\ref{SPT_beam_fig})
and Section (\ref{is_diel})). For comparison, the level of
spurious effects produced by a good feed featured by low
cross-polarization (e.g. - 40 dB of BaR-SPOrt
\cite{spurious,carretti_2}), even though not optimised to minimize
such a systematic, is $\sim$ - 25 dB, as represented by the
horizontal line traced across the right-plot of Fig.
(\ref{SPT_beam_fig}). Such a level matches the requirements for
CMBP experiments. The systematics generated by flat dielectric
slabs prevail on that produced by good feeds for beams greater
than $\sim$ 15$^{\circ}$, thus requiring either thickness
optimization analysis or the choice survey to control spurious
polarizations.
\\
The estimates of the depolarization effects introduced by the
dielectrics are shown in Fig.
(\ref{depolarization_isotropic_fig}).
\\
Close to the axis we find that $\langle
\tilde{D}_{Q}^{T}(d,\nu,\theta^{\; \prime}) \rangle_{\nu} \sim
\langle \tilde{D}_{Q}^{T}(d,\nu,\theta) \rangle_{\nu}$, then it is
possible to estimate Eq. (\ref{eta_q_t_q0_const}) approximating
$\langle \tilde{D}_{Q}^{T}(d,\nu,\theta^{\; \prime})
\rangle_{\nu}$ with $\langle \tilde{D}_{Q}^{T}(d,\nu,\theta)
\rangle_{\nu}$. Thanks to the selected thickness and materials,
the loss of the polarized signal is marginal ($\sim 10^{-3}$). It
is worth noting that the thickness $d_{0}$ maximizing the
transmittance also minimizes the depolarization (see the magnified
frame of right panel in Fig.
(\ref{depolarization_isotropic_fig})).
\\
Finally, the leakages are estimated. In a conservative approach,
the $Q_{U^{0}}^{T}$ leakage term (Eq. (\ref{q_u0})) can be
estimated as:
\begin{eqnarray}\label{leakage_q_u0}
    Q_{U^{0}}^{T}(d) & < & |\Delta \! {\rm{U}^{0}}(\nu,\theta,\beta)|_{\rm MAX(\nu,\theta,\beta)}
                    \times f_{T,M_{33}^{T}}(d) \\
    f_{T,M_{33}^{T}}(d) & = & \frac{2 \pi}{\Omega_{A}^{\rm FF}} \int_{0}^{\pi/2} [ \langle {T}(d,\nu,\theta)
                    \rangle_{\nu} - \langle {M_{33}^{T}}(d,\nu,\theta) \rangle_{\nu} ]\; P_{n}^{\rm FF}(\theta) \sin(\theta)
                    \; \rm{d}\theta \label{leakage_q_u0_2}
\end{eqnarray}
The approximation holds because the term ($\langle T \rangle_{\nu}
- \langle M_{33}^{T} \rangle_{\nu}$) has a monotone growing trend
for $\theta_{i}$ comparable with the BaR-SPOrt beam ($\sim
7^{\circ}$), as shown in top-left panel of Fig.
(\ref{leakage_isotropic_fig}). For the selected thickness and
materials, the maximum leakage from $U^{0}$ to $Q$ in the
30.4--33.6 GHz band is about 0.07$\times|\Delta \! U^{0}|_{\rm
MAX}$. A more rigorous computation, which takes into account the
beam pattern, provides the result shown in
Fig.~(\ref{leakage_isotropic_fig}) (bottom-left panel). The
minimum leakage is realized with the thickness $d_{0}$ for which
the values drop down to negligible values ($10^{-8} \div 10^{-9}$
with respect to $|\Delta \! U^{0}|_{\rm MAX}$). Same results hold
also for $U_{Q^{0}}^{T}$.
\\
Similarly, the $Q_{V^{0}}^{T}$ term (Eq. (\ref{q_v0})) can be
estimated as:
\begin{eqnarray}\label{leakage_q_v0}
    Q_{V^{0}}^{T}(d) & < & |\Delta \! {\rm{V}^{0}}(\nu,\theta,\beta)|_{\rm MAX(\nu,\theta,\beta)}
                    \times f_{\!L^{\!T}}(d) \\
    f_{\!L^{\!T}}(d) & = & \frac{2 \pi}{\Omega_{A}^{\rm FF}} \int_{0}^{\pi/2} \langle |L^{T}(d,\nu,\theta)|
                    \rangle_{\nu} \; P_{n}^{\rm FF}(\theta) \sin(\theta)
                    \; \rm{d}\theta \label{leakage_q_v0_2}
\end{eqnarray}
where the considerations done for ($\langle T \rangle_{\nu} -
\langle M_{33}^{T} \rangle_{\nu}$) can be extended to $\langle
|L^{T}|\rangle_{\nu}$. Teflon, HDPE and Polypropylene introduce in
the 30.4--33.6 GHz band a 0.16 maximum leakage of $V^{0}$ into $Q$
and $U$. If such a function is smoothed by the beam pattern, then,
for the selected thickness $d_{0}$ which also here minimize the
effect, the leakage becomes negligible ($10^{-4} \div 10^{-5}$
with respect to $|\Delta \! {\rm{V}^{0}}|_{\rm{MAX}}$).
\\
Again, among the selected materials, Teflon introduces the lowest
leakage and depolarization effects.
\\
Note that the depolarization and leakage in transmission are
signal losses which can be recovered by an overall instrument
calibration.
\section{Conclusion}
In this work we presented the systematic effects introduced by a
flat slab of isotropic dielectric.
\\
We presented an overall analysis of the interaction between
electromagnetic radiation and isotropic dielectric at microwave
frequencies, by analyzing transmittance, reflectance, absorptance,
spurious polarization, leakage and depolarization by means of the
Mueller formalism.
\\
The important result is that spurious polarization, and leakage
between the Stokes parameters are produced even by optically
isotropic dielectrics, but only when they are thermally
inhomogeneous or the incident radiation is anisotropic.
In particular, it has been provided an estimate of the expected
systematic effects introduced by Teflon, Polypropylene and HDPE,
together with algorithms for their thickness optimization to
minimize the effects.
\\
Measurements of dielectric constant and electric tangent loss of
Teflon, HDPE, Polypropylene and Nylon have been provided between
27 GHz and 37 GHz at 300 K of physical temperature. The Teflon
sample analyzed is featured by the lowest $\langle \epsilon_r
\rangle_{\nu} \sim$ 2.04 and $\langle \tan(\delta_e) \rangle_{\nu}
\sim$ 1.6 $\times$ 10$^{-4}$ averaged in the 30.4--33.6 GHz band.
Moreover, we found that no optical anisotropy at level of 1$\%$
has been measured from our Teflon sample about the index of
refraction.
\\
The analysis shows that Teflon, among the selected materials, is
the best material in the investigated band which minimizes the
systematic effects.
\\
As discussed in Section (\ref{application}), the optimal thickness
to maximize transmission ($\sim 0.999$) and reduce emission ($\sim
0.00054$) is $d = 0.33$ cm. The maximum thermal noise introduced
by this slab is 162 mK at a physical temperature of T$_{{\rm ph}}
= 300$ K. This thickness, which minimizes also the transmitted
spurious polarization ($< 2.6 \times 10^{-5}$, thus producing in
emission a spurious level $< 6 \times 10^{-7}$), the leakage ($<
10^{-8}$ from $Q^{0}$ to $U$, or $U^{0}$ to $Q$, $\sim$ 5~$\times$
10$^{-5}$ from $V^{0}$ to $Q$ or $U$) and the depolarization
($\sim$ 1.3 $\times$ 10$^{-3}$), corresponds to $d = \lambda/2n$.
Broadly speaking, in the approximation of low absorbing material
and high feed horn directivity, such a thickness maximizes the
transmission and reduces all the other effects.
\\
We have also shown that for dielectrics in far field regime, the
transmitted spurious polarization prevails on the one produced by
good feeds (cross-polarization $\sim$ -- 40 dB) for beams greater
than $\sim$ 15$^{\circ}$, thus showing the need of either
thickness optimization analysis or the choice survey of flat
dielectrics to control spurious polarizations. The position of
such thicknesses, which set maxima and minima of the spurious
response, depends on the beam adopted.
\section*{Acknowledgments}
Authors wish to thanks Renzo Nesti and Vincenzo Natale for useful
discussion, and the anonymous referee for the useful comments and
the encouragement to improve the paper. This work is inserted in
the BaR-SPOrt program, an experiment aimed at detecting the CMBP,
which is funded by ASI (Italian Space Agency).
\newpage
\begin{table}   [!h]
        \caption{Geometrical sizes of the samples: $a$ is the height,
                $b$ is the width and $l$ is the length. The dimensions of the WR28
                standard rectangular waveguide are: $(a,b)=(3.556,7.112)$ mm.}
    \begin{center}
        \begin{tabular}{||c|c|c|c||}
        \hline
                            & $ a \; [\textrm{mm}]$ & $ b \; [\textrm{mm}]$ & $ l \; [\textrm{mm}]$ \\
        \hline
        \hline
        Teflon 1-cut        & 3.51 $\pm$ 0.05 & 7.10 $\pm$ 0.03 & 59.90 $\pm$ 0.07     \\
        \hline
        Teflon 2-cut        & 3.39 $\pm$ 0.09 & 7.08 $\pm$ 0.03 & 59.85 $\pm$ 0.04     \\
        \hline
        HDPE                & 3.54 $\pm$ 0.05 & 6.93 $\pm$ 0.09 & 60.08 $\pm$ 0.10     \\
        \hline
        Polypropylene       & 3.58 $\pm$ 0.03 & 7.09 $\pm$ 0.13 & 59.98 $\pm$ 0.08     \\
        \hline
        Nylon               & 3.53 $\pm$ 0.05 & 7.11 $\pm$ 0.07 & 60.21 $\pm$ 0.13     \\
        \hline
        \end{tabular}
        \label{materials_size_table}
    \end{center}
\end{table}
\begin{table}   [!h]
        \caption{Complex dielectric constant and index of refraction with their maximum errors for: HDPE, Teflon,
                Polypropylene and Nylon. The values provided are in-band averages (30.4--33.6 GHz).}
    \begin{center}
        \begin{tabular}{||c|c|c|c|c||}
        \hline
                            & $ \langle \epsilon_{r} \rangle_\nu$ & $\langle \tan(\delta_{e})\rangle_\nu \;\;
                            [10^{-4}]$ & $ \langle n \rangle_\nu$ & $\langle \kappa \rangle_\nu \;\; [10^{-4}]$ \\
        \hline
        \hline
        HDPE                & 2.32 $\pm$ 0.03 & 1.7 $\pm$ 0.7 & 1.523 $\pm$ 0.010 & 1.30 $\pm$ 0.54      \\
        \hline
        Teflon              & 2.04 $\pm$ 0.02 & 1.6 $\pm$ 0.4 & 1.428 $\pm$ 0.007 & 1.14 $\pm$ 0.29      \\
        \hline
        Polypropylene       & 2.24 $\pm$ 0.03 & 4.6 $\pm$ 1.8 & 1.497 $\pm$ 0.010 & 3.44 $\pm$ 1.37      \\
        \hline
        Nylon               & 3.00 $\pm$ 0.02 & 111.0 $\pm$ 1.7 & 1.732 $\pm$ 0.006 & 96.13 $\pm$ 1.79     \\
        \hline
        \end{tabular}
        \label{optical_constant_risultati}
    \end{center}
\end{table}
\begin{table}   [!h]
        \caption{In band (30.4--33.6 GHz) variation, with respect to their average
                value, of $\epsilon_{r}$ and $\tan(\delta_{e})$
                for HDPE, Teflon, Polypropylene and Nylon.}
    \begin{center}
        \begin{tabular}{||c|c|c||}
        \hline
                            & $ \eta_{\epsilon_{r}}$ & $ \eta_{\tan(\delta_{e})}$ \\
        \hline
        \hline
        HDPE                & 0.2 $\%$   & 30 $\%$              \\
        \hline
        Teflon              & 0.1 $\%$   & 25 $\%$              \\
        \hline
        Polypropylene       & 0.1 $\%$   & 10 $\%$              \\
        \hline
        Nylon               & 0.2 $\%$   & 2 $\%$              \\
        \hline
        \end{tabular}
        \label{in_band_var_risultati}
    \end{center}
\end{table}
\begin{table}   [!h]
        \caption{Bibliographic data of complex dielectric constants for HDPE, Teflon and Polypropylene.
                The last column refers to bibliography. No datum is available for Nylon.}
    \begin{center}
        \begin{tabular}{||c|c|c|c|c||}
        \hline
                      & $\nu$ [GHz] &  $\epsilon_{r}$  &  $\tan (\delta_{e})$  &  Reference     \\
        \hline
        \hline
        HDPE          & 35.26       &  2.359    &   0.00017  & \cite{jones}              \\
        \hline
        Teflon        & 34.54       &  1.95     &   0.00005    & \cite{jones}       \\
        \hline
        Polypropylene & 35          &  2.254    &   0.00015  &  \cite{afsar}      \\
        \hline
        \end{tabular}
        \label{opt_constant_ref}
    \end{center}
\end{table}
\begin{table}   [!h]
        \caption{In-band average (30.4--33.6 GHz) with their maximum errors
                of the complex dielectric constant and index of refraction for
                both 1-cut and 2-cut Teflon. The parameters
                $\langle \tan(\delta_{e})\rangle_\nu$ and
                $\langle \kappa \rangle_\nu$ are in $[10^{-4}]$ units. Last two
                columns report the in band variation of $\bf{\epsilon_r}$.}
    \begin{center}
        \begin{tabular}{||c|c|c|c|c|c|c||}
        \hline
                            & $ \langle \epsilon_{r} \rangle_\nu$ & $\langle \tan(\delta_{e})\rangle_\nu
                            $ & $ \langle n \rangle_\nu$ & $\langle \kappa \rangle_\nu $
                            & $\eta_{\epsilon_r}$ & $\eta_{\tan(\delta_e)}$ \\
        \hline
        \hline
        1-cut        & 2.04 $\pm$ 0.02 & 1.6 $\pm$ 0.4 & 1.428 $\pm$ 0.007 & 1.14 $\pm$ 0.29
                            & $\pm$ 0.1 $\%$ & $\pm$ 25 $\%$    \\
        \hline
        2-cut        & 2.01 $\pm$ 0.02 & 2.2 $\pm$ 0.7 & 1.418 $\pm$ 0.007 & 1.56 $\pm$ 0.50
                            & $\pm$ 0.1 $\%$ & $\pm$ 10 $\%$    \\
        \hline
        \end{tabular}
        \label{Teflon_anis_table}
    \end{center}
\end{table}
\begin{table}   [!bht]
    \caption{Conservative estimates of the noise temperature
            introduced by the flat dielectric slab evaluated
            at T$_{\rm{ph}}^{{\rm MAX}} = 300$ K. The thickness $d_{0}$ is 0.33 cm
            for Teflon, 0.31 cm for HDPE and Polypropylene.}
    \begin{center}
        \begin{tabular}{||c|c|c|c||}
        \hline
                    & $\langle \varepsilon(d,\nu,\theta) \rangle_{\nu}|_{{\rm MAX}(\theta)} $
                    & ${\rm T}_{A}^{{\rm \varepsilon}}(d)/{\rm T}_{{\rm ph}}^{{\rm MAX}} $ & $\textrm{T}_{\rm noise}$ [mK] \\
        \hline
        \hline
        Teflon         & 6.1 $\times$ 10$^{-4}$ & 5.4 $\times$ 10$^{-4}$  &  162 \\
        \hline
        HDPE           & 6.6 $\times$ 10$^{-4}$ & 5.9 $\times$ 10$^{-4}$  &  177 \\
        \hline
        Polypropylene  & 17.0 $\times$ 10$^{-4}$ & 15.4 $\times$ 10$^{-4}$&  462 \\
        \hline
        \end{tabular}
        \label{noise_temp_dielectrics}
    \end{center}
\end{table}
\begin{table}   [!h]
    \caption{Upper limits of the spurious polarization as from Eq.
            (\ref{over_est}) introduced by HDPE, Teflon and Polypropylene isotropic slab.
                The thickness adopted is $d_{0}$. $\Delta \! \rm{T}_{b}^{{\rm MAX}} = 100 \mu$K and
                $\Delta \! \rm{T}_{\rm{ph}}^{{\rm MAX}} = 1$ K. (The same for U).}
    \begin{center}
        \begin{tabular}{||c|c|c|c|c|c||}
        \hline
                      &  $S\!P^{T}_{{\rm MAX}}$  & $Q_{S\!P}^{T}|_{{\rm MAX}}$  & $S\!P^{\varepsilon}_{{\rm MAX}}$
                      &  $Q_{S\!P}^{\varepsilon}|_{{\rm MAX}}$   &  $Q_{S\!P}|_{{\rm MAX}}$  \\
        \hline
        \hline
        HDPE          &  0.240  & 24 $\mu$K  &  0.00015   & 0.15 mK   &    0.174 mK     \\
        \hline
        Teflon        &  0.215  &  $\sim$ 22 $\mu$K  &  0.00012   & 0.12 mK   &    0.142 mK    \\
        \hline
        Polypropylene &  0.235  & $\sim$ 24 $\mu$K  &  0.00038   & 0.38 mK   &    0.404 mK     \\
        \hline
        \end{tabular}
        \label{sp_finestra_isotropa}
    \end{center}
\end{table}
\begin{table}   [!h]
    \caption{Upper limits of the spurious polarization, as from
            Eq. (\ref{major_stokes_30_bis})-(\ref{major_spe}), introduced by HDPE, Teflon
            and Polypropylene isotropic slab by taking into account the BaR-SPOrt feed.
            The thickness adopted is $d_{0}$. $\Delta \! \rm{T}_{b}^{{\rm MAX}} =~100 \mu$K
            and $\Delta \! \rm{T}_{\rm{ph}}^{{\rm MAX}} = 1$ K. (The same for U).}
    \begin{center}
        \begin{tabular}{||c|c|c|c|c|c||}
        \hline
                      &
$|Q_{S\!P}^{T}|/\Delta \!{\rm T\!}_{\rm b}^{{\rm MAX}}$ &
$Q_{S\!P}^{T}|_{{\rm MAX}}$  & $|Q_{S\!P}^{\varepsilon}|/\Delta\!
{\rm T\!}_{{\rm ph}}^{{\rm MAX}}$  &
$Q_{S\!P}^{\varepsilon}|_{{\rm MAX}}$
& $Q_{S\!P}|_{{\rm MAX}}$  \\
        \hline
        \hline
HDPE          &  3.1 $\times$ 10$^{-5}$ & 0.0031 $\mu$K  &  8.8 $\times$ 10$^{-7}$  & 0.88 $\mu$K   & $\sim$ 0.9 $\mu$K     \\
        \hline
Teflon        &  2.6 $\times$ 10$^{-5}$ & 0.0026 $\mu$K  &  6.0 $\times$ 10$^{-7}$  & 0.60 $\mu$K   & $\sim$ 0.6 $\mu$K    \\
        \hline
Polypropylene &  3.7 $\times$ 10$^{-5}$ & 0.0037 $\mu$K  &  18.8 $\times$ 10$^{-7}$ & 1.88 $\mu$K   & $\sim$ 1.9 $\mu$K     \\
        \hline
        \end{tabular}
        \label{sp_finestra_isotropa_2}
    \end{center}
\end{table}
\clearpage
\vspace{10 mm}
\begin{figure} [h]
    \centerline{\includegraphics[width=4in,height=3in]{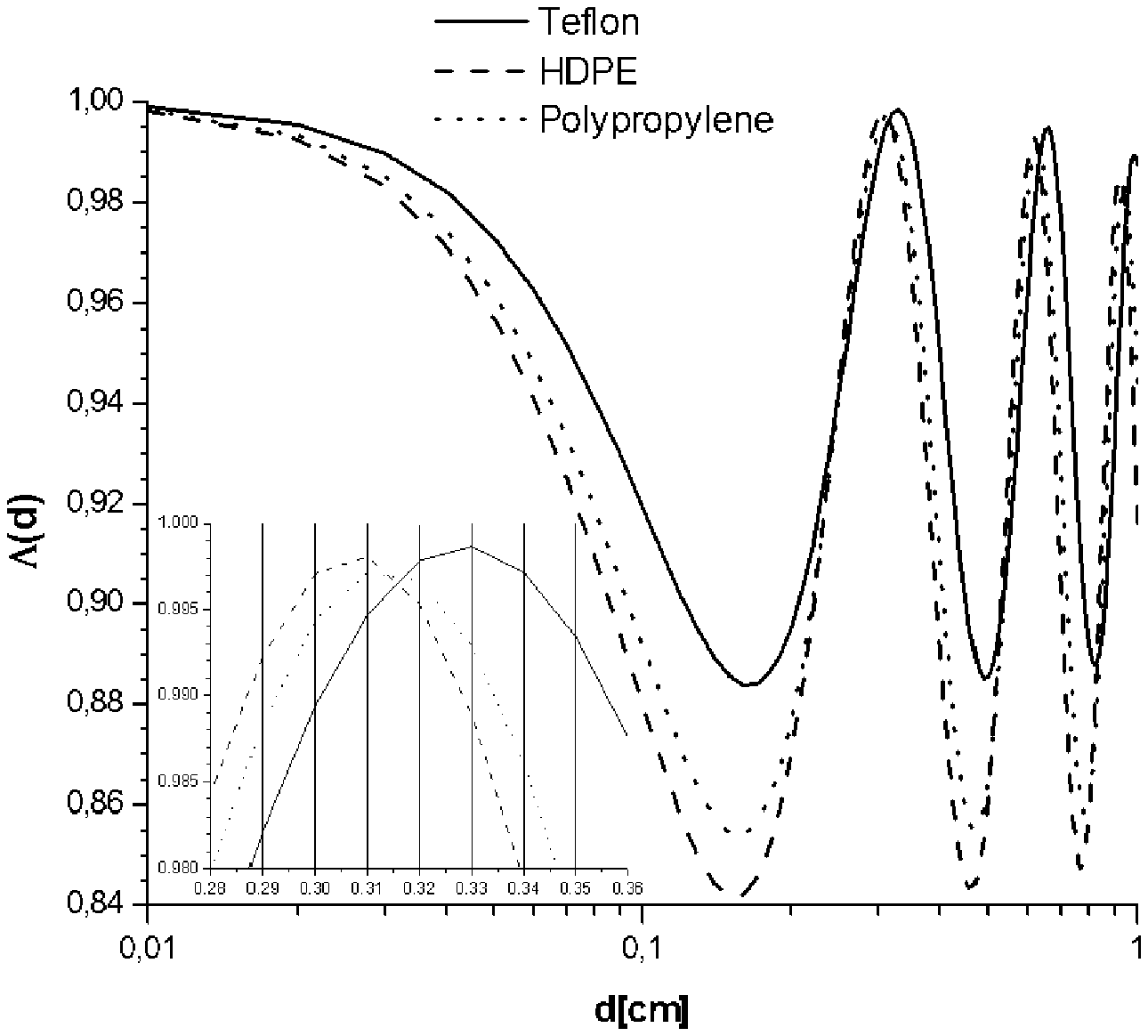}}
\end{figure}
\begin{figure} [h]
    \centerline{\includegraphics[width=3.5in,height=2.5in]{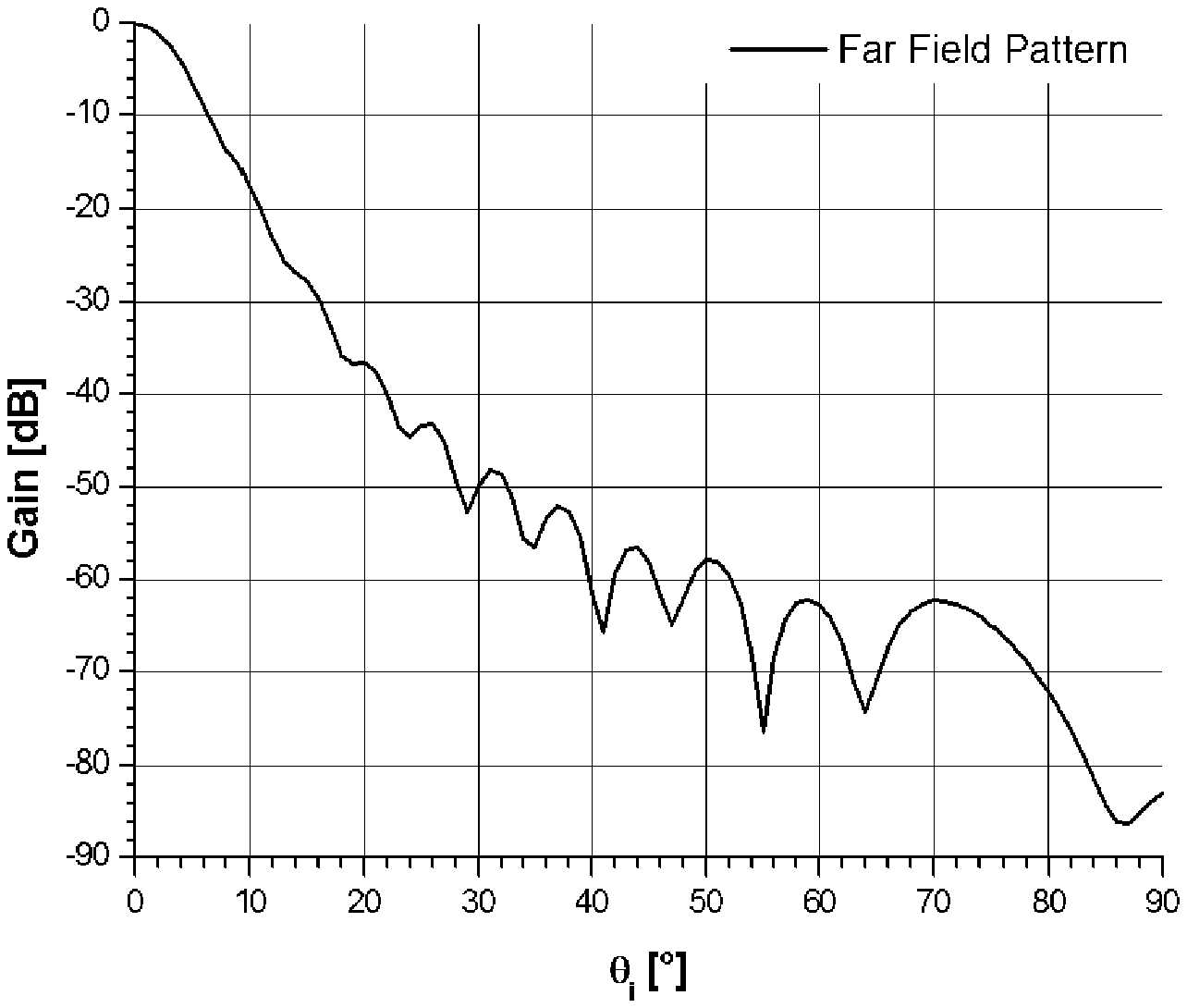}}
\end{figure}
\clearpage
\begin{figure} [h]
    \centerline{\includegraphics[width=3.5in,height=2.5in]{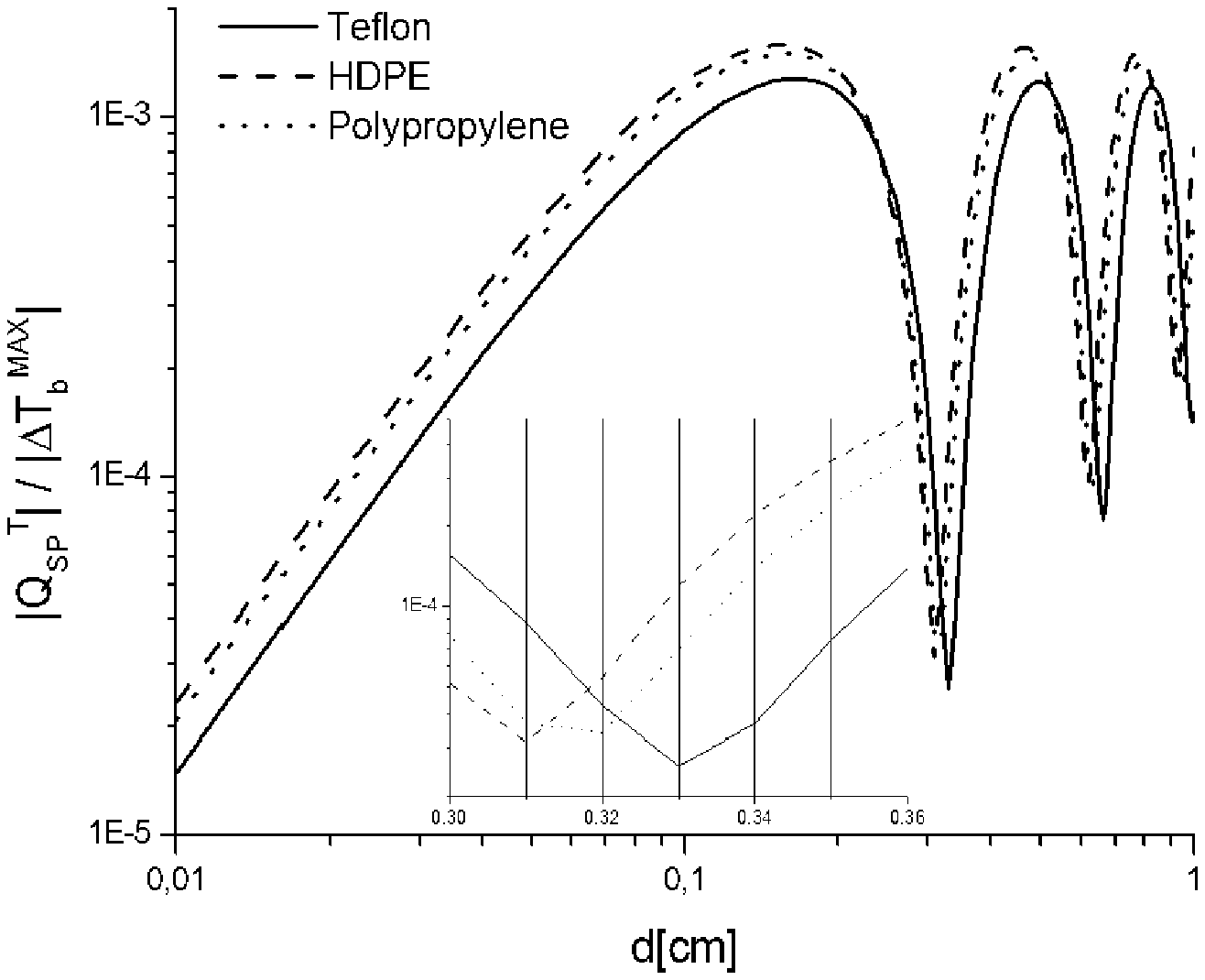}}
\end{figure}
\clearpage
\begin{figure} [h]
    \centerline{\includegraphics[width=3.3in,height=2.4in]{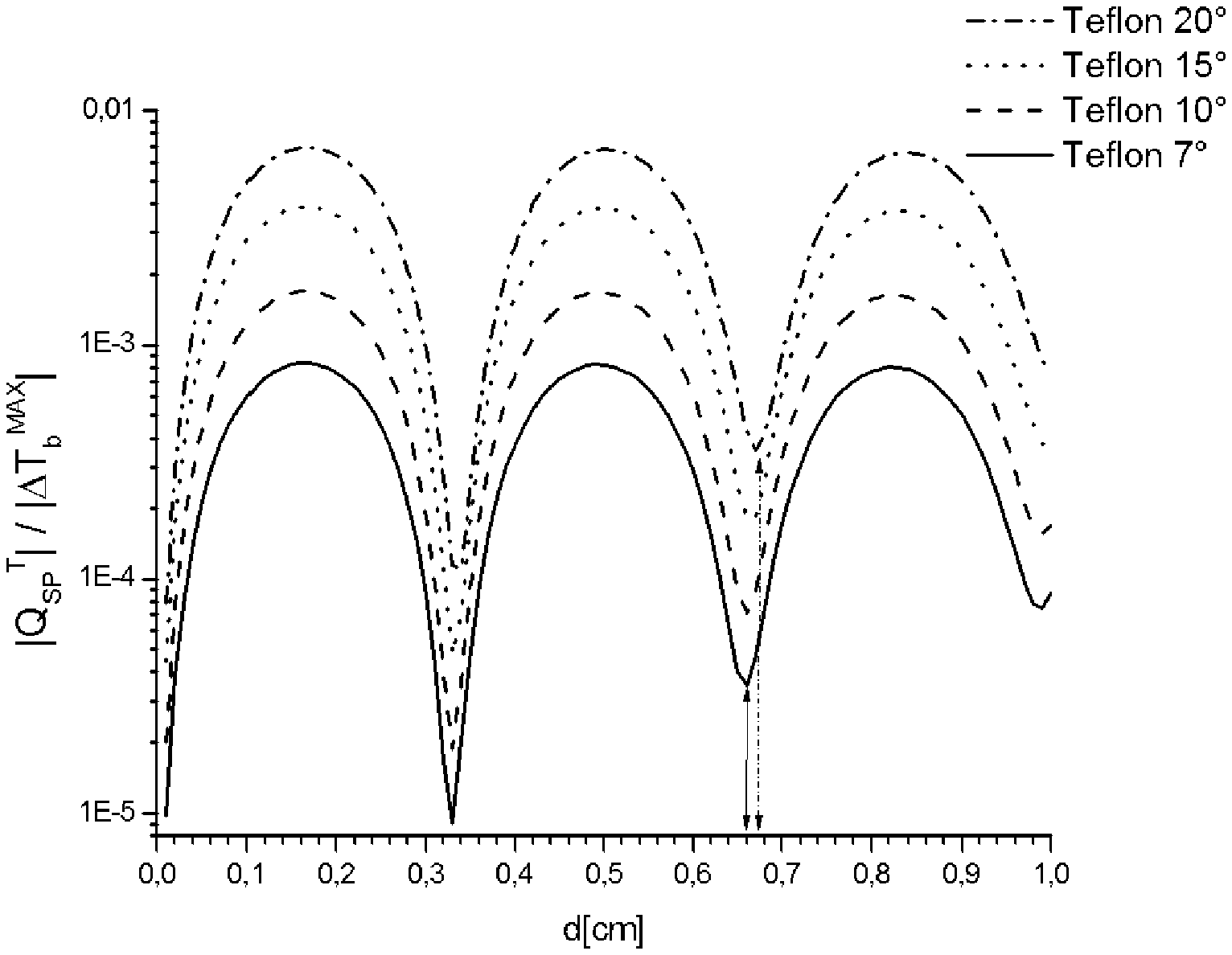}}
\end{figure}
\clearpage
\begin{figure} [h]
    \centerline{\includegraphics[width=3.4in,height=2.4in]{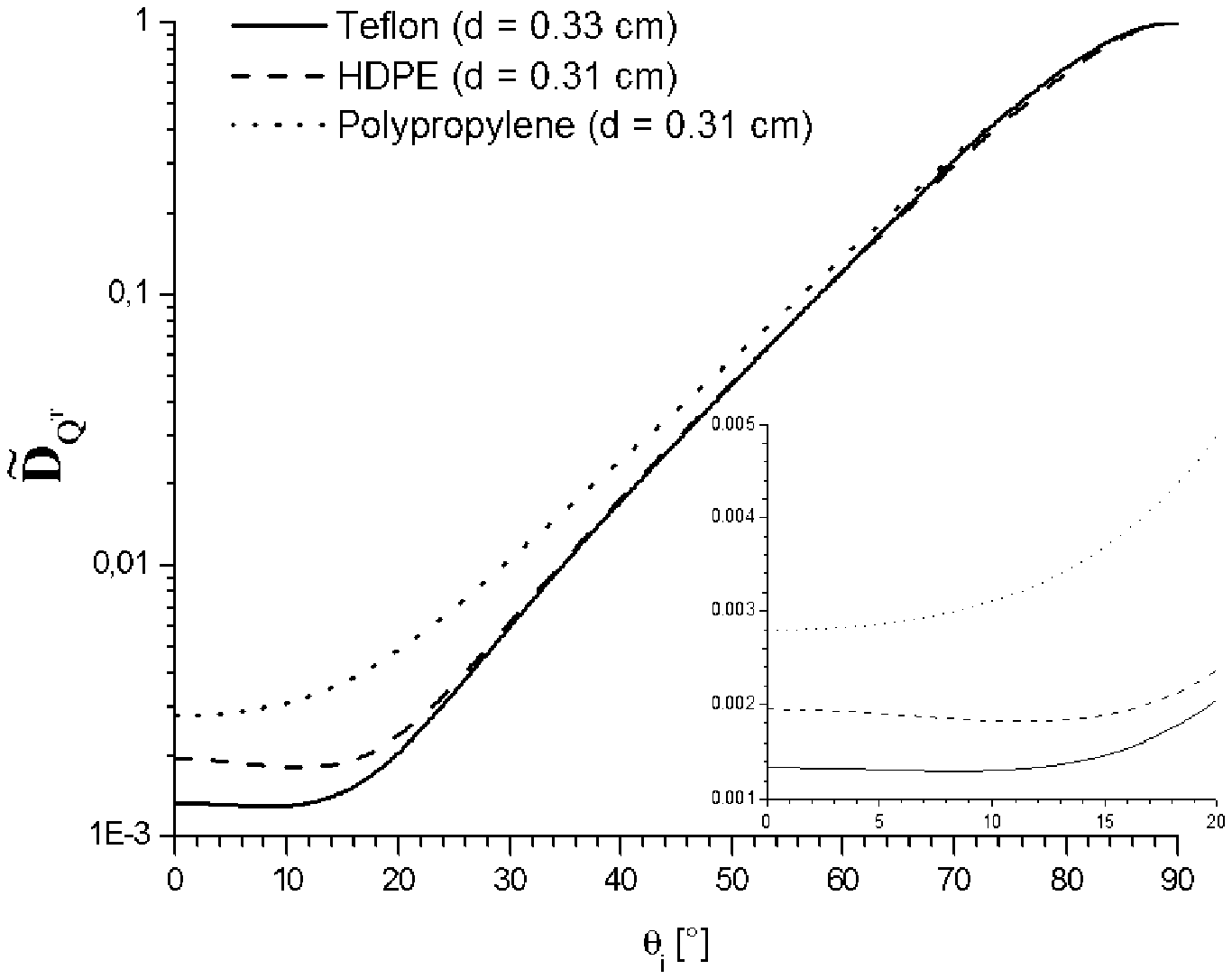}}
\end{figure}
\clearpage
\end{document}